\documentclass[reprint,nofootinbib,amsmath,amssymb,showkeys,aps]{revtex4-1}

\usepackage[utf8]{inputenc}
\usepackage{amsmath,amssymb}
\usepackage{graphicx}
\usepackage{float}
\usepackage{subcaption}
\usepackage{tabularx}
\usepackage{bm}
\usepackage{natbib}
\usepackage{comment}
\usepackage[normalem]{ulem}

\usepackage{hyperref}
\hypersetup{
    colorlinks=true,
    linkcolor=blue,
    filecolor=blue,      
    urlcolor=blue,
    citecolor=blue
    }

\usepackage[dvipsnames]{xcolor}
  
\usepackage[T1]{fontenc} 
\usepackage{braket}

\definecolor{lcol}{HTML}{006699}

\usepackage[nameinlink,noabbrev]{cleveref}
\Crefname{equation}{Eq.}{Eqs.}
\Crefname{section}{Sec.}{Sects.}
\Crefname{figure}{Fig.}{Figs.}

\defcitealias{IST:paper1}{EC19}

\begin{document}

\title{Constraining constant and tomographic coupled dark energy\\ with low-redshift and high-redshift probes}

\author{Lisa W. K. Goh$^1$}\email{lisa.goh@cea.fr}
\author{Adri\`a G\'omez-Valent$^{2,3}$}\email{agvalent@roma2.infn.it}
\author{Valeria Pettorino$^1$}\email{valeria.pettorino@cea.fr}
\author{Martin Kilbinger$^1$}

\affiliation{$^1$ Université Paris-Saclay, Université Paris Cité, CEA, CNRS, Astrophysique, Instrumentation et Modélisation Paris-Saclay, 91191 Gif-sur-Yvette, France}
\affiliation{$^2$ Dipartimento di Fisica, Università di Roma Tor Vergata, via della Ricerca Scientifica 1, 00133 Roma, Italy}
\affiliation{$^3$ INFN, Sezione di Roma 2, Università di Roma Tor Vergata, via della Ricerca Scientifica 1, 00133 Roma, Italy}

\begin{abstract}
We consider coupled dark energy (CDE) cosmologies, where dark matter particles feel a force stronger than gravity, due to the fifth force mediated by a scalar field which plays the role of dark energy. We perform for the first time a tomographic analysis of coupled dark energy, where the coupling strength is parametrised and constrained in different redshift bins. This allows us to verify which data can better constrain the strength of the coupling and how large the coupling can be at different epochs. First, we employ cosmic microwave background data from $\textit{Planck}$, the Atacama Cosmology Telescope (ACT) and South Pole Telescope (SPT), showing the impact of different choices that can be done in combining these datasets. Then, we use a range of low redshift probes to test CDE cosmologies, both for a constant and for a tomographic coupling. In particular, we use for the first time data from weak lensing (the KiDS-1000 survey), galaxy clustering (BOSS survey), and their combination, including 3x2pt galaxy-galaxy lensing cross-correlation data.
We do not find evidence for non-zero coupling, either for a constant or tomographic case. A non-zero coupling is however still in agreement with current data. For CMB and background datasets, a tomographic coupling allows for $\beta$ values up to one order of magnitude larger than in previous works, in particular at $z < 1$.
The use of 3x2pt analysis then becomes important to constrain $\beta$ at low redshifts, even when coupling is allowed to vary: for 3x2pt we find, at $0.5 < z < 1$, $\beta = 0.018^{+0.007}_{-0.011}$, comparable to what CMB and background datasets would give for a constant coupling. This makes upcoming galaxy surveys potentially powerful probes to test CDE models at low redshifts. 
We find a smaller tension in $H_0$ and $S_8$ when the coupling is allowed to vary, although this is rather due to an increase in their uncertainties. We also see that our model is penalised by a Bayesian ratio model comparison with respect to $\Lambda$CDM, which is still favoured by current data.

\end{abstract}

\maketitle
\section{Introduction}\label{intro}
In the past decade, numerous cosmological surveys and advancements in observational techniques have allowed researchers to place percent-level constraints on cosmological parameters, bringing us closer to understanding the nature of our Universe. So far, the $\Lambda$CDM concordance model has proven to be the most favoured; yet, pertinent questions remain unanswered. Two concerns are often discussed; for a review, see \cite{Abdalla:2022yfr} and references therein. Firstly, the Hubble tension: why is the value of the Hubble constant measured to be ${\sim}5\sigma$ lower using high-redshift probes like the cosmic microwave background (CMB) \cite{Planck:2018vyg}, compared to using low redshift probes such as distance ladder measurements \cite{Riess:2021jrx}? Secondly, the $S_8$ tension: why is the amplitude of clustering, measured through the cosmological parameter $S_8=\sigma_8\sqrt{\Omega_\mathrm{m}/0.3}$, calculated by \textit{Planck} \cite{Planck:2018vyg}, higher than that measured by weak lensing surveys \cite{DES:2017myr,KiDS:2020suj} by ${\sim}2.5\sigma$? If systematics are under control, one possibility would be to question the sufficiency of our current standard model and whether introducing new physics is necessary to explain these problems. In this paper, we shall focus on one class of $\Lambda$CDM model extensions, known as the coupled dark energy (CDE) model.

Past works have proposed such a model where there exists a coupling between dark matter particles of varying mass, mediated by a scalar field that plays the role of dark energy. Such models have been first hypothesised as a solution to the coincidence problem \cite{Amendola:1999er,Wetterich:1994bg}, and have subsequently been studied in a number of papers \cite{1111.1404,1207.3293,Pettorino:2008ez, Pettorino:2013oxa,2003PhRvD..68b3514A,2000PhRvD..62d3511A, Gomez-Valent:2020mqn,Archidiacono:2022iuu,Gomez-Valent:2022bku}. CDE models are among the $\Lambda$CDM model extensions which are still in agreement with current data \cite{Gomez-Valent:2020mqn} and which may help with solving the Hubble tension (\cite{Pettorino:2013oxa,DiValentino:2019ffd}; they were, however, proposed before the Hubble tension became significant, and independently of it. In these works, the coupling is quantified by a coupling strength parameter, assumed to take on a constant value for simplicity. Varying couplings have also been considered in literature, such as in the case of a vacuum dark energy interaction with dark matter particles \cite{Martinelli:2019dau,Hogg:2020rdp}, between dark matter particles themselves 
 (see eg. \cite{Barrow:2006hia,Baldi:2008ay,Caldera-Cabral:2008yyo,Baldi:2010vv}) and analogously, between neutrino particles \cite{Wetterich:2007kr,Pettorino:2010bv}. 

In this work, we wish to consider the following questions: what is the impact of a varying coupling on the main cosmological observables? Up to what extent can we constrain this evolution with current data? Is the coupling strength similarly bounded in all the stages of cosmic expansion? Would we be able to achieve a similar, or even better, degree of alleviation of the $H_0$ tension with a varying coupling? To this end, we first propose a new phenomenological `tomographic coupling' parameterisation, in which we allow for different amplitudes of the coupling at different epochs. We then bin the coupling strength parameter in different redshifts. In doing so, we also update constraints on a constant coupling, using real data from weak lensing, galaxy clustering and for the first time the galaxy-galaxy lensing 3x2 pt observable, including cross-correlation terms; furthermore, we use updated baryonic acoustic oscillations (BAO) and Type Ia supernovae (SNe1a) data, as well as the updated SH0ES prior. 

This paper is organised as follows: in Sec.~\ref{sec:CDE} we give an overview of the theoretical framework behind the CDE model and define our tomographic coupling model. In Sec.~\ref{sec:Data} we illustrate the different datasets we use to constrain our model and motivate our choices. We explain our methodology and define the various binning schemes we will be studying in Sec.~\ref{sec:Method}; we present our results in Sec.~\ref{sec:Results}. We conduct a model comparison in Sec.~\ref{sec:model-comparison}. Finally, we draw our conclusions in Sec.~\ref{sec:conclusions}.
\section{Coupled Dark Energy}\label{sec:CDE}
We consider CDE cosmologies in which the mass of dark matter particles is dependent on a scalar field, $m_{\rm cdm}=m_{\rm cdm}(\phi)$, and assumed to be fermionic. We refer to \cite{Pettorino:2008ez, Baldi:2008ay} for a detailed description of the formalism and here recall only the main equations that describe these cosmological models. We consider the following Lagrangian density for the dark sector, 
\begin{equation}
\mathcal{L}_{\rm dark}=-\partial_\mu\phi\partial^\mu\phi-V(\phi)-m_{\rm cdm}(\phi)\bar{\psi}\psi+\mathcal{L}_{\rm kin}[\psi]\,,
\end{equation} 
with $\mathcal{L}_{\rm kin}[\psi]$ the dark matter (DM) kinetic term and $V(\phi)$ the potential of the scalar field $\phi$. The latter is assumed to be responsible for the late-time acceleration of the Universe, i.e. it plays the role of dark energy, and mediates an interaction between the DM particles, which feel a fifth force in addition to the standard gravitational pull. The baryonic sector is not coupled to $\phi$, so the model automatically fulfils the very stringent local constraints on the weak equivalence principle and screened fifth forces \cite{Will:2014kxa,Elder:2019yyp,Berge:2017ovy, Brax:2018zfb}.

We present in Sec. \ref{sec:cosmoEq} the main cosmological equations of the model, allowing for general forms of the coupling function and the scalar field potential. We subsequently focus in Sec. \ref{sec:tomobeta} on a particular class of coupled DE scenarios with a tomographic coupling and comment on some of its most relevant phenomenological features.
\subsection{Cosmological Equations}\label{sec:cosmoEq}
The interaction between $\phi$ and the dark matter component leads to the following two covariant conservation equations: 

\begin{equation}\label{eq:covCons}
    \nabla^\mu T_{\mu\nu}^{\phi}=\kappa\beta T^{\rm cdm}\nabla_\nu\phi\quad ;\quad  \nabla^\mu T_{\mu\nu}^{\rm cdm}=-\kappa\beta T^{\rm cdm}\nabla_\nu\phi\,,    
\end{equation}
with $T_{\mu\nu}^{\phi}$ and $T_{\mu\nu}^{\rm cdm}$ the DE and DM energy-momentum tensors  respectively, and $T^{\rm cdm}=g^{\mu\nu}T_{\mu\nu}^{\rm cdm}$. The total energy-momentum tensor of the dark sector is conserved; i.e. $\nabla^\mu (T_{\mu\nu}^{\phi}+T_{\mu\nu}^{\rm cdm})=0$, and in general the dimensionless coupling $\beta$ can be a function of $\phi$. In this paper, we use the natural unit $c=1$ and denote the inverse of the reduced Planck mass as $\kappa=\sqrt{8\pi G}$.

We assume the Cosmological Principle and a flat Universe, so at the background level we use the flat Friedmann-Lema\^itre-Robertson-Walker metric, and consider perturbations on top of it. In this paper, we present the perturbed equations in the synchronous gauge, since it is the one employed in our modified version of \texttt{CLASS}, see Sec. \ref{sec:tomobeta}. For the equations in the Newtonian gauge, cf. \cite{Pettorino:2008ez,Baldi:2008ay}.

In the synchronous gauge, the metric reads 

\begin{equation}\label{eq:FLRW}
\mathrm{d}s^2=a^2(\tau)[-\mathrm{d}\tau^2+\{\delta_{ij}+h_{ij}(\tau,\vec{x})\}\mathrm{d}x^i\mathrm{d}x^j]\,,
\end{equation}
with $\tau$ the conformal time and $\vec{x}$ the spatial comoving vector. The two scalar degrees of freedom of $h_{ij}$, 

\begin{equation}
h_{ij} = \int \mathrm{d}^3k\, e^{-\mathrm{i}\overrightarrow{k}\cdot\overrightarrow{x}}\left[h\,\hat{k}_i\hat{k}_j+6\eta\left(\hat{k}_i\hat{k}_j-\frac{\delta_{ij}}{3}\right)\right]\,,
\end{equation}
are contained in its trace $h(\tau,\vec{k})$ and $\eta(\tau,\vec{k})$, with $\vec{k}$ the comoving wavenumber \cite{Ma:1995ey}. 

\subsubsection{Background Equations}\label{sec:background}

In the coupled DE model under consideration the energy density and pressure associated with $\phi$ read: 

\begin{equation}
\rho_\phi=\frac{(\phi^\prime)^2}{2a^2}+V(\phi)\quad ;\quad p_\phi=\frac{(\phi^\prime)^2}{2a^2}-V(\phi)\,,
\end{equation} 
with the primes denoting derivatives with respect to conformal time. Substituting \eqref{eq:FLRW} at leading order into Eq. \eqref{eq:covCons} we obtain the modified Klein-Gordon equation: 

\begin{equation}\label{eq:KG}
\phi^{\prime\prime}+2\mathcal{H}\phi^{\prime}+a^2\frac{\partial V}{\partial\phi}=\kappa \beta a^2\rho_{\rm cdm}\,,
\end{equation}
which governs the cosmic background evolution of $\phi$, and the modified equation for the dark matter density,

\begin{equation}\label{eq:consrhoDM}
\rho^\prime_{\rm cdm}+3\mathcal{H}\rho_{\rm cdm}=-\kappa\beta\rho_{\rm cdm}\phi^{\prime}\,,
\end{equation}
where $\mathcal{H}\equiv a^{\prime}/a$ and $\rho_{\rm cdm}$ is the mean (background) DM energy density. These equations do not depend on the derivatives of $\beta$ and, hence, they are exactly the same as the ones found in the $\beta=$constant case, see e.g. \cite{Gomez-Valent:2020mqn}. The right-hand side of \eqref{eq:KG} and \eqref{eq:consrhoDM} tells us that a non-zero coupling only has a non-zero impact on the background dynamics when the DM fraction is non-negligible, i.e. in the matter-dominated epoch onwards. 

The scalar field potential can also accelerate $\phi$ at different epochs of the cosmic expansion, depending on its shape. In this study, though, we use a constant potential, i.e.
\begin{equation}
    V(\phi)=V_0\,
\end{equation}
which is in charge of the late-time acceleration of the universe and has no impact on the scalar field dynamics at high redshifts. For the sake of generality, in Sec. \ref{sec:perturbations} we still provide the expressions considering a non-constant potential.
\subsubsection{Perturbation Equations}\label{sec:perturbations}
We define the density contrast of a particular species $i$ as $\delta_i\equiv\delta\rho_i/\rho_i$, with $\delta\rho_i$ the corresponding density perturbation, and $\delta\phi$ the perturbation of $\phi$. The linearly perturbed conservation equation of the scalar field \eqref{eq:covCons} leads, in momentum space, to

\begin{multline}\label{eq:pertKG}
    \delta\phi^{\prime\prime}+2\mathcal{H}\delta\phi^\prime+\left(k^2+a^2\frac{\partial^2V}{\partial\phi^2}\right)\delta\phi+\frac{h^\prime}{2}\phi^\prime=\\
    \kappa\rho_{\rm cdm}a^2\left(\beta\delta_{\rm cdm}+\frac{\partial\beta}{\partial\phi}\delta\phi\right)\,,
\end{multline}
whereas the evolution of the DM density contrast and velocity divergence takes the following form, respectively:

\begin{equation}\label{eq:pertConsDens}
\delta^\prime_{\rm cdm}=-\theta_{\rm cdm}-\frac{h^\prime}{2}-\kappa\beta\delta\phi^\prime-\kappa\frac{\partial\beta}{\partial\phi}\phi^\prime\delta\phi\,
\end{equation}

\begin{equation}\label{eq:pertThetaPrime}
\theta^\prime_{\rm cdm}=(-\mathcal{H}+\kappa\beta\phi^\prime)\theta_{\rm cdm}-k^2\kappa\beta\delta\phi\ \,\, .
\end{equation}
Notice that Eqs. \eqref{eq:pertKG} and \eqref{eq:pertConsDens} contain a term that depends on the derivative of $\beta$, which is absent when $\beta$ is constant. The equations obtained from the perturbed Einstein equations are the standard $\Lambda$CDM ones, but receive, of course, the contribution of the DE density and pressure perturbations,

\begin{equation}\label{eq:pertphi}
\delta\rho_\phi=\frac{\phi^\prime\delta\phi^\prime}{a^2}+\frac{\partial V}{\partial\phi}\delta\phi\quad ;\quad \delta p_\phi=\frac{\phi^\prime\delta\phi^\prime}{a^2}-\frac{\partial V}{\partial\phi}\delta\phi
\end{equation}
and its velocity divergence, 

\begin{equation}
\theta_\phi=k^2\delta\phi/\phi^\prime\,.
\end{equation}
At subhorizon scales, i.e. for $k\gg\mathcal{H}$, under a reasonably slow evolution of $\beta$ satisfying $\partial\beta/\partial(\kappa\phi)\leq\mathcal{O}(1)$ and the massless DE scalar field we are considering in this study, the following equation for the DM density contrast applies:

\begin{equation}\label{eq:densityContrastDM}                  
\begin{aligned}
\delta_{\rm cdm}^{\prime\prime}+&\left[\mathcal{H}-\beta\kappa\phi^\prime\right] \delta^\prime_{\rm cdm}\\
& -\frac{\kappa^2a^2}{2}\left[\rho_b\delta_b+\rho_{\rm cdm}\delta_{\rm cdm}(1+2\beta^2)\right]=0\,.
\end{aligned}
\end{equation}
Regardless of the sign of $\beta$, the last term of Eq. \eqref{eq:densityContrastDM} is enhanced by the interaction in the dark sector. On the other hand, the friction term decreases if the sign of $\beta$ does not change throughout cosmic history. These two effects lead to a change in the clustering of matter in the Universe as compared to  $\Lambda$CDM, for fixed initial conditions \cite{Pettorino:2008ez,Baldi:2008ay}. For a constant $\beta$ the DM density contrast evolves as $\delta_{\rm cdm}\sim a^{1+2\beta^2}$ during the matter-dominated era.

\begin{figure}[t!]
\begin{center}
 \includegraphics[width=3.2in, height=2.5in]{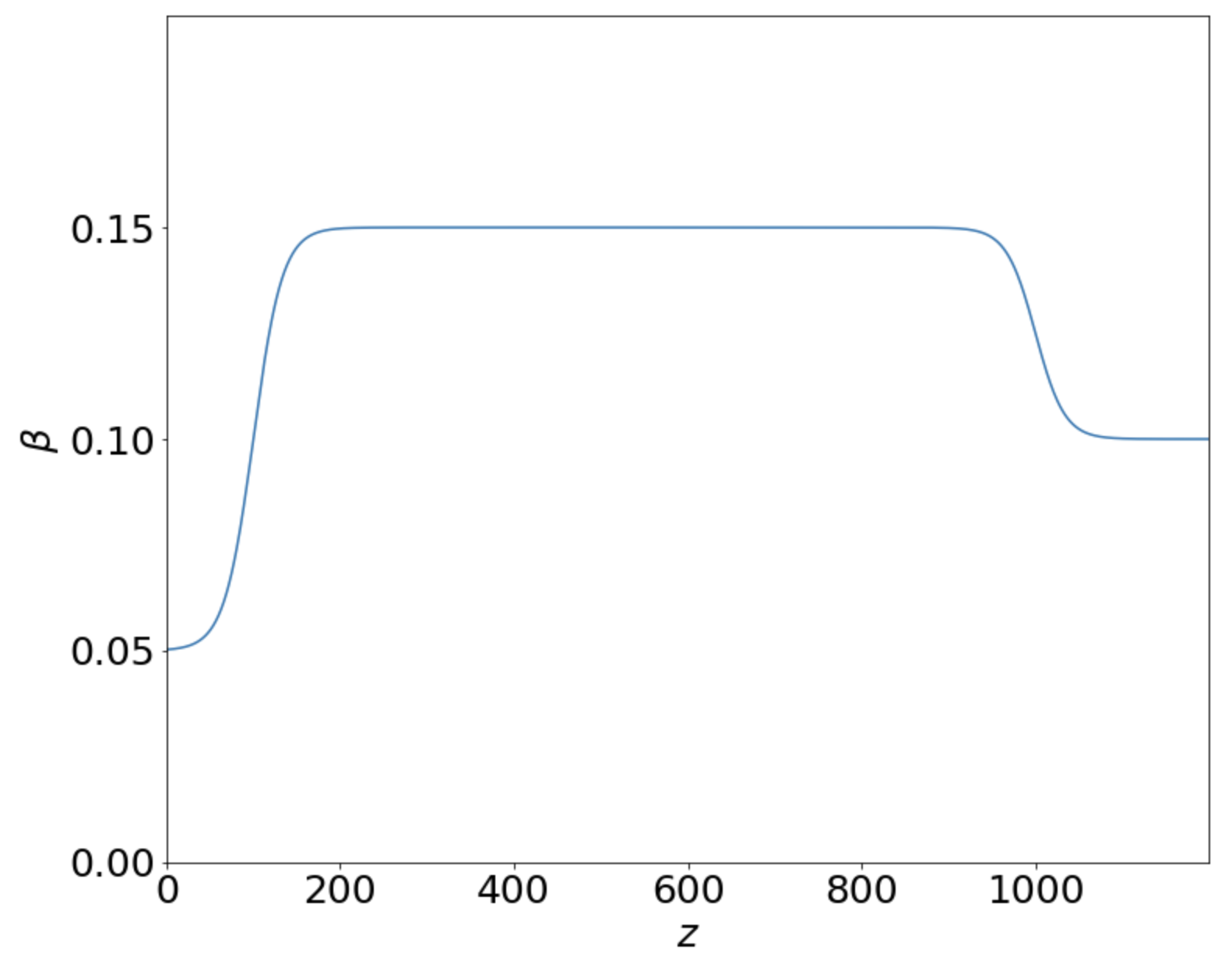}
 \caption{Example of $\beta(z)$ with three bins, using $s_i=0.03$ for $i=1,2$ and arbitrarily chosen values of $\beta_1=0.05$ ($z\lesssim10^2$), $\beta_2=0.15$ ($10^2\lesssim z\lesssim 10^3$), $\beta_3=0.10$ ($z\gtrsim 10^3$).}\label{fig:beta}
\end{center}
\end{figure}

Another point worth noting is the possibility of the sign of $\beta$ changing in time. This could lead to periods of cosmic history with a larger friction term, which would slow down the evolution of $\delta_{\rm cdm}$. However, here we keep $\beta\geq 0$ for simplicity. 
\subsection{Tomographic Coupling}\label{sec:tomobeta}
Here we explain the theoretical formalism of our tomographic CDE model. Instead of exploring specific functional forms of $\beta=\beta(\phi)$ we opt to make use of a more general, phenomenological, expression for the coupling, written directly in terms of the redshift, i.e. $\beta=\beta(z)$, which is easy to interpret and visualise.

\begin{figure*}[ht!]
\begin{center}
    \begin{subfigure}[b]{0.45\linewidth}
            \centering
            \includegraphics[trim = 30 35 47 100, clip,width=1.05\linewidth]{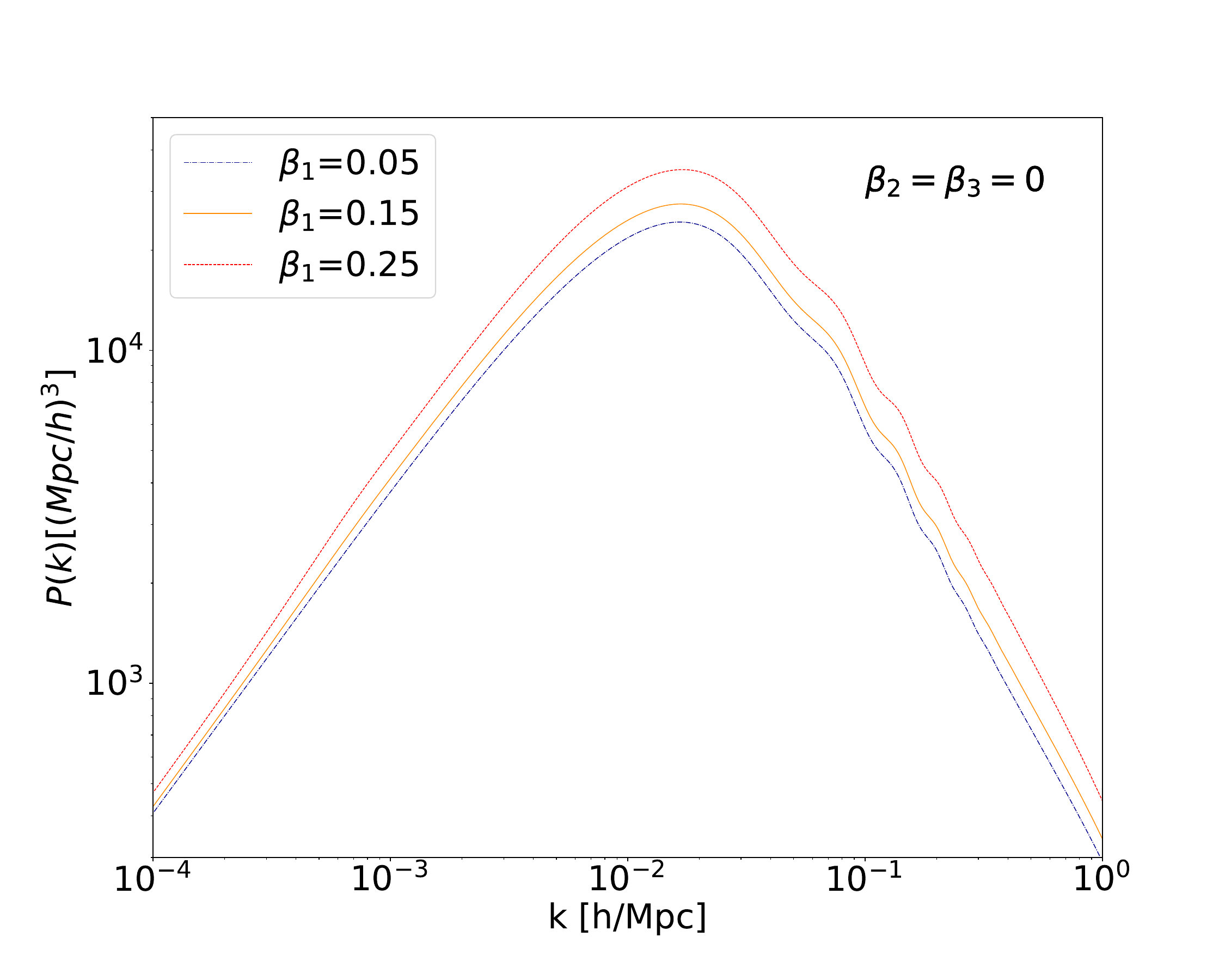}
        \end{subfigure}
        \begin{subfigure}[b]{0.45\linewidth}  
            \centering 
            \includegraphics[trim = 30 35 70 80, clip,width=1.05\linewidth]{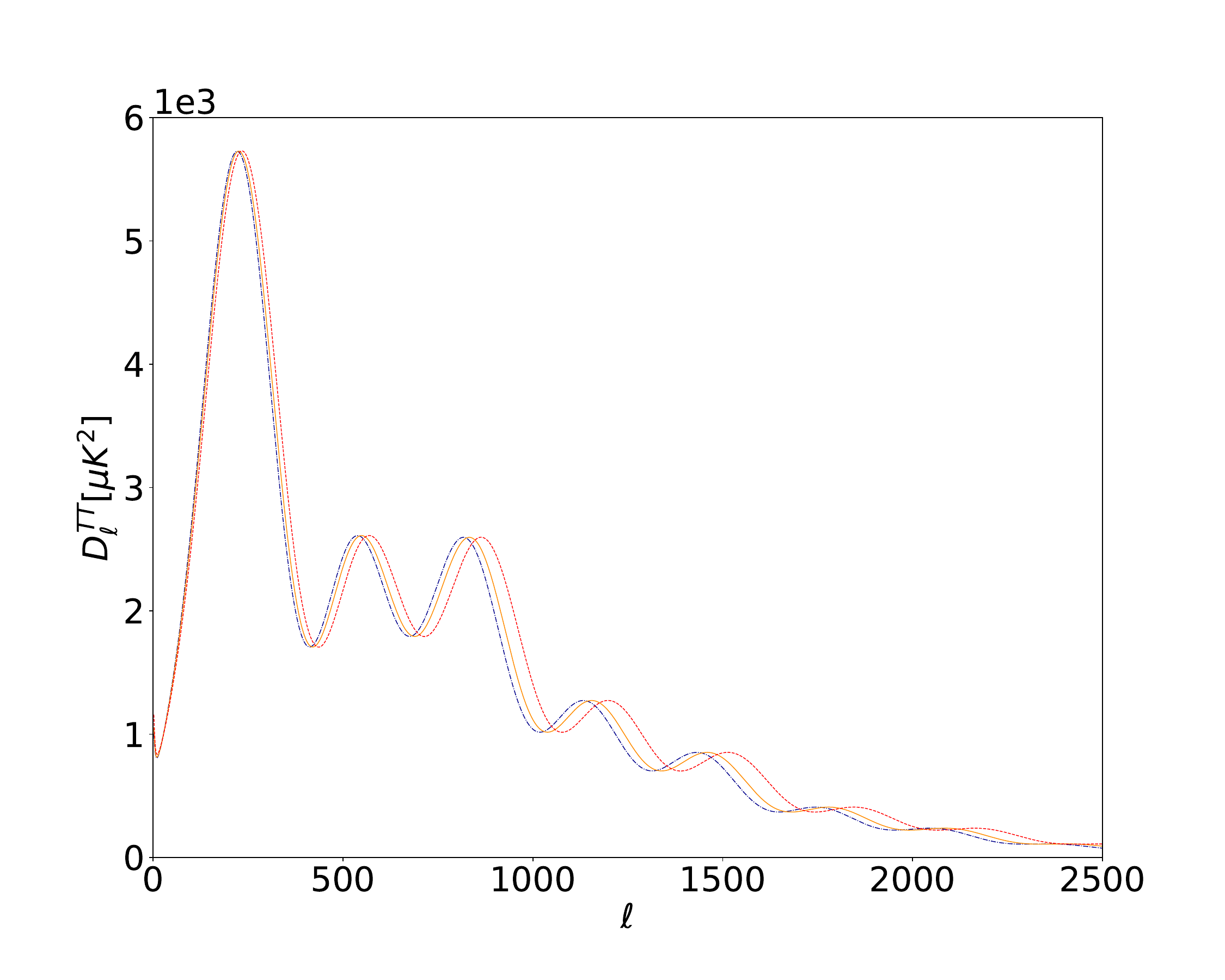}
        \end{subfigure}
        \vskip\baselineskip
        \begin{subfigure}[b]{0.45\linewidth}   
            \centering 
            \includegraphics[trim = 30 35 47 100, clip,width=1.05\linewidth]{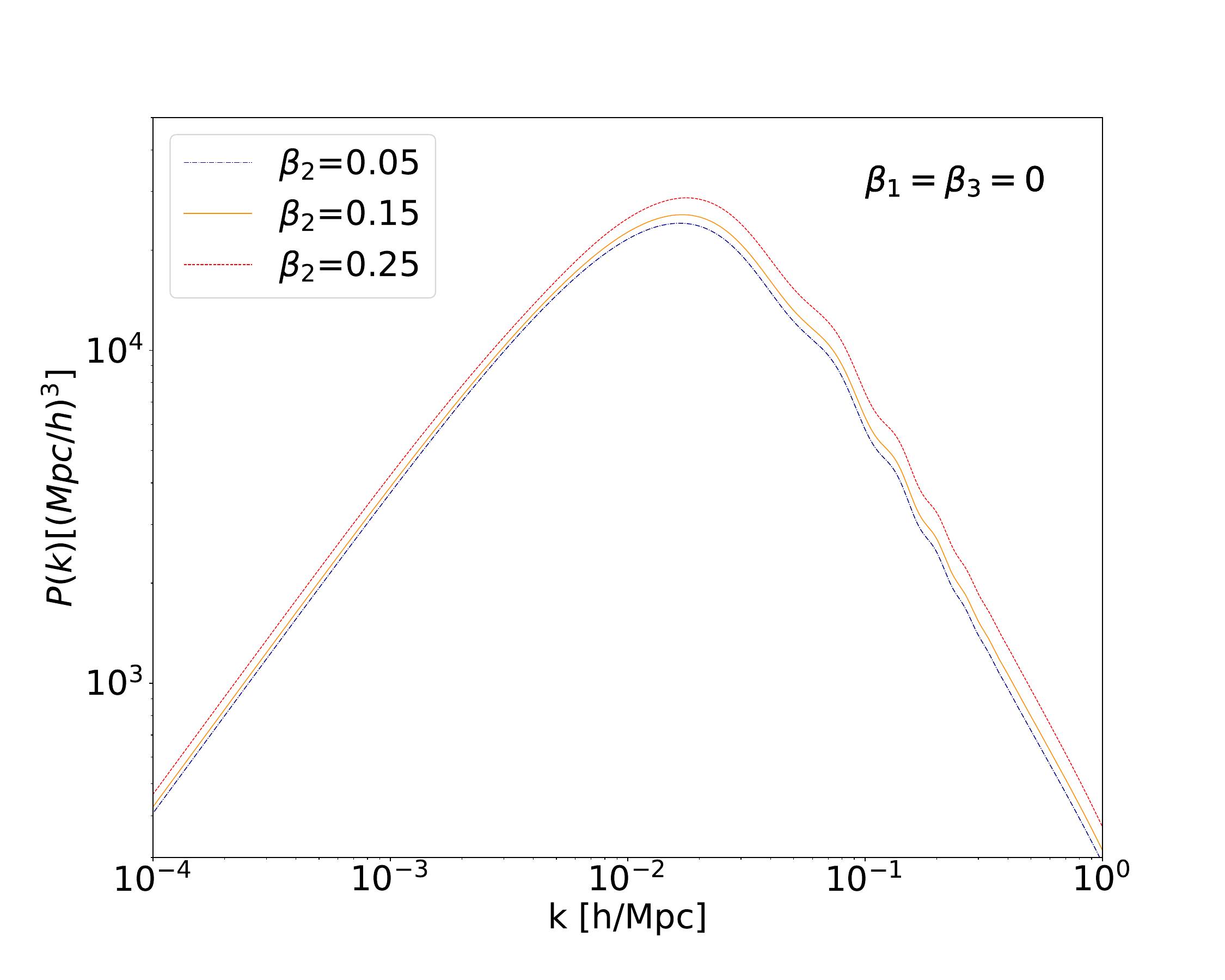}
        \end{subfigure}
        \begin{subfigure}[b]{0.45\linewidth}   
            \centering 
            \includegraphics[trim = 30 35 70 80, clip,width=1.05\linewidth]{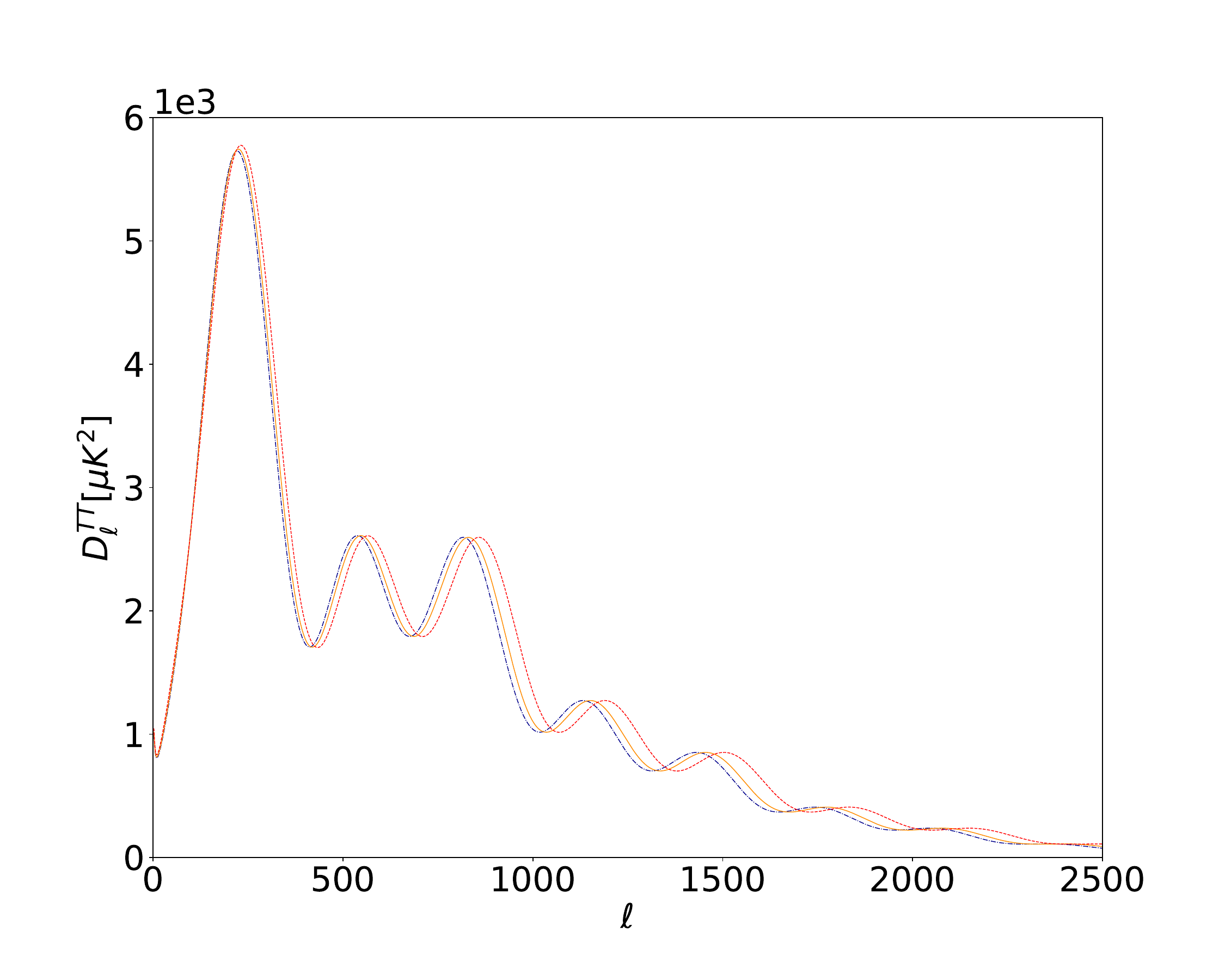}
        \end{subfigure}
        \vskip\baselineskip
            \begin{subfigure}[b]{0.45\linewidth}
            \centering
            \includegraphics[trim = 30 35 47 100, clip,width=1.05\linewidth]{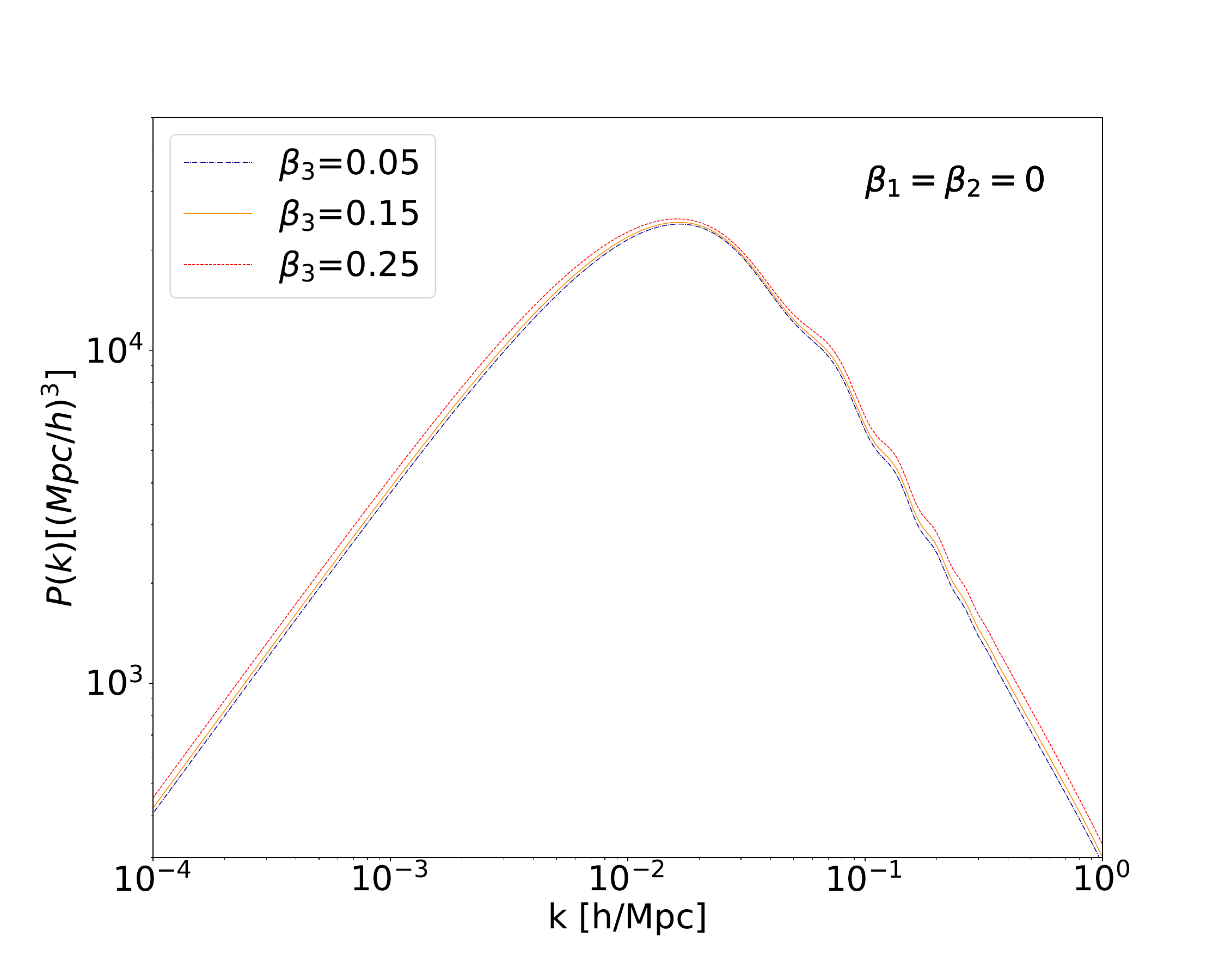}
        \end{subfigure}
        \begin{subfigure}[b]{0.45\linewidth}  
            \centering 
            \includegraphics[trim = 30 35 70 80, clip,width=1.05\linewidth]{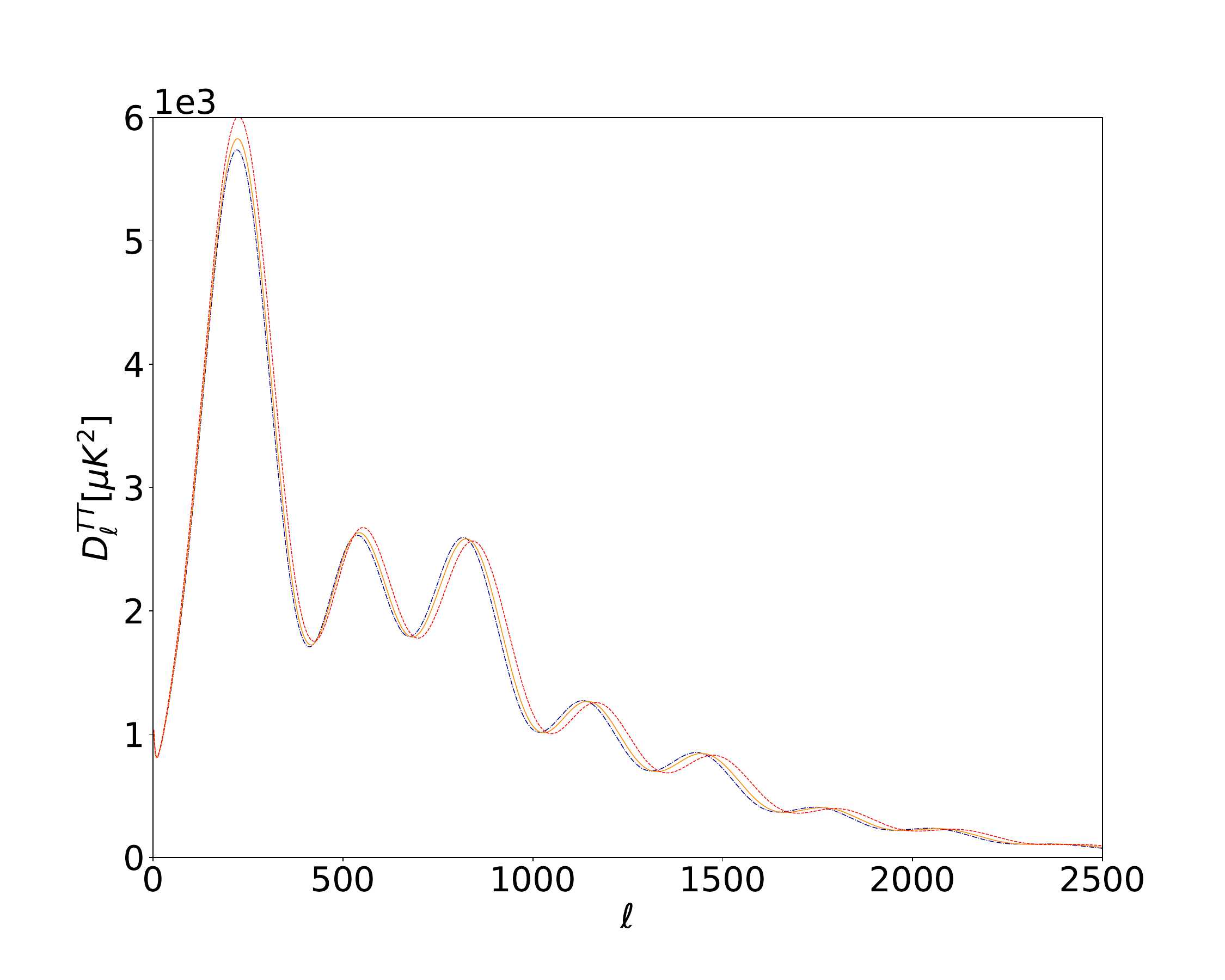}
        \end{subfigure}

        \caption{Matter power spectra $P(k)$ (left column) and CMB temperature spectra $D_\ell^{TT}$ (right column) obtained with different values of $\beta_1$ (top row), $\beta_2$ (middle row) and $\beta_3$ (bottom row), while setting the other $\beta_i$'s to zero. This is to show the impact of a non-null coupling at each bin. We recall the reader that the bin edges are defined as $z=\{0,100,1000\}$. We use in all cases the same primordial power spectrum, $\tau_\mathrm{reio}$, $\omega_\mathrm{b}$, $V_0$ and the initial value of CDM energy density as in the {\it Planck} TT+TE+EE $\Lambda$CDM best-fit cosmology.} \label{fig:3binSpectra}
\end{center}
\end{figure*}

We consider a smooth function built from hyperbolic tangents,
\begin{equation}\label{eq:binbeta}
\beta(z) = \frac{\beta_1+\beta_n}{2}+\frac{1}{2}\sum_{i=1}^{n-1}(\beta_{i+1}-\beta_{i})\tanh[s_i(z-z_i)]\,,
\end{equation}
where $n$ is the number of redshift bins and $z_i$ is the upper limit of the $i_{\rm th}$ bin. We can easily express the derivatives of $\beta$ appearing in Eq. \eqref{eq:pertKG} and \eqref{eq:pertConsDens} as follows,
\begin{equation}
\frac{\partial\beta}{\partial \phi} = -\frac{\partial\beta}{\partial z}\frac{\mathcal{H}}{a\phi^\prime}\,,
\end{equation}     
with
\begin{equation}
\frac{\partial\beta}{\partial z}=\frac{1}{2}\sum_{i=1}^{n-1}\frac{s_i(\beta_{i+1}-\beta_i)}{\cosh^2[s_i(z-z_i)]}\,.
\end{equation}
The constants $s_i$ are smoothing factors that control the steepness of $\beta(z)$ when transitioning between tomographic bins. They have to be chosen such that they allow to reach the values of $\beta_i$ in the $i_{\rm th}$ bin $\forall{i}\in[1,n]$. In Fig.~\ref{fig:beta} we plot an example form that $\beta(z)$ could take, assuming a 3-bin model with bin edges $z=[0,100,1000]$.

We have extended the modified version of \texttt{CLASS} \cite{2011JCAP...07..034B} employed in \cite{Gomez-Valent:2020mqn,Gomez-Valent:2022bku} to accommodate a redshift-dependent coupling $\beta(z)$\footnote{Our modified version of \texttt{CLASS} is publicly available at the following link: \url{https://github.com/LisaGoh/CDE}, together with some notes to facilitate its use.}, see Sec. \ref{sec:Method} for details. In Fig.~\ref{fig:3binSpectra} we plot the current linear matter and CMB temperature power spectra obtained when we switch on the coupling in only one bin of the 3-bin scenario to understand the effects that an early-, mid- and late-time coupling can have on these cosmological observables. To ensure a fair comparison between
the resulting curves, we fix in all the cases studied
in this figure the same values of the standard
cosmological parameters $n_\mathrm{s}$, $A_\mathrm{s}$, $\tau_\mathrm{reio}$, $\omega_\mathrm{b}$, the cosmological constant $V_0$, and the initial value of the DM energy density (at $z=10^{14}$) following {\it Planck} $\Lambda$CDM best-fit
cosmology. Thus, the differences between these curves are only due to the activation of the different coupling windows and the value they take on.

\begin{figure}
    \centering
    \includegraphics[width=1.05\linewidth]{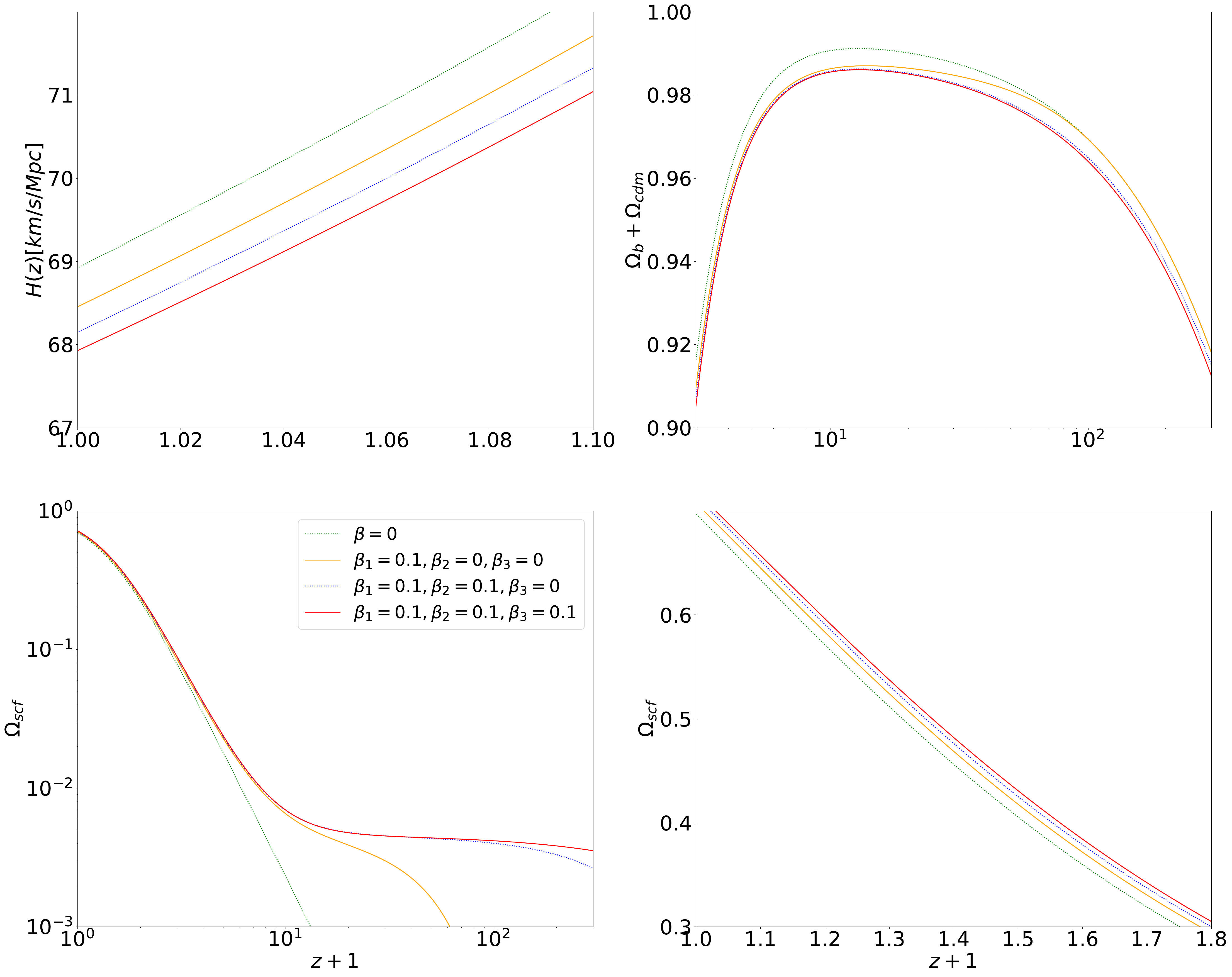}
    \caption{We plot the evolution of $H(z)$ and the values of $\Omega_{b}(z)+\Omega_{\rm cdm}(z)$ and $\Omega_{\phi}(z)$ when we subsequently switch on the coupling strength at increasing redshifts by activating the corresponding coefficient of $\beta$ from $0$ to $0.1$. The green dotted line with $\beta=0$ represents the $\Lambda$CDM model. The plot of $\Omega_\phi(z)$ at the bottom right is a zoom-in of the bottom left plot, in the window $0.1 \leq z \leq 0.8$, which allows us to grasp the differences between the various curves. }
    \label{fig:varying_beta}
\end{figure}

Fig.~\ref{fig:3binSpectra} shows that the matter power spectrum is most sensitive to the coupling in the first two tomographic bins and, more conspicuously, in the first one, i.e., at $z<\mathcal{O}(10^2)$. This is because this is the period of the cosmic expansion at which the scaling solution is typically reached by the scalar field (cf. the lower left plot of Fig. \ref{fig:varying_beta}) and, hence, when the fifth force leaves a larger imprint on the structure formation processes. In any case, we see that a non-null coupling generates an increase in the amplitude of matter fluctuations regardless of the activation bin, as already discussed in Sec.~\ref{sec:perturbations}. Regarding the CMB TT power spectra, we observe that the peaks are enhanced and move to larger multipoles for larger values of the coupling. The enhancement is bigger when the coupling is activated in the pre-recombination epoch, since it decreases the ratio $\rho_{\rm cdm}/\rho_{\rm b}$, whereas the shift to larger multipoles is more important when the coupling is switched on at $z<1000$, in the first two bins. This is because for fixed initial conditions, we have a lower amount of dark matter at the decoupling time and at very small redshifts. The lowering of $H(z)$ leads to a larger sound horizon at the baryon-drag epoch $r_s$ and a larger angular diameter distance to the last scattering surface, but the late-time effect is more important and this is why the angle subtended in the sky by $r_s$ decreases. Notice that when the coupling is only activated in the third bin, i.e., at $z\gtrsim1000$, the shifts induced in the power spectra are in general much smaller than those found when the coupling is switched on after decoupling, as the DM fraction in the pre-recombination epoch is smaller and so are the source terms in the background conservation equations for DE and DM, i.e. the right-hand side of Eqs. \eqref{eq:KG} and \eqref{eq:consrhoDM}.

Subsequently, in Fig.~\ref{fig:varying_beta}, we show the evolution of several background quantities obtained again with the 3-bin tomographic model, fixing the initial conditions in all cases as in Fig.~\ref{fig:3binSpectra}. In particular, we highlight the effect that initialising the coupling $\beta$ at different redshifts has on the late-time evolution of $H(z)$ and the DE and DM energy fractions $\Omega_{i}(z)$. Since $\rho_{\rm cdm}$ decreases with increasing $\beta$, as seen from Eq. \eqref{eq:consrhoDM}, the activation of the coupling leads to lower values of $H_0$. 

\section{Data}\label{sec:Data}
Here we list the datasets we use in this work to constrain the CDE model described in Sec.~\ref{sec:CDE}. With respect to \cite{Gomez-Valent:2020mqn} we update our BAO, SNe1a and SH0ES datasets, and add new probes for the first time, namely: the Atacama Cosmology Telescope (ACT) and the South Pole Telescope (SPT) for the CMB, with more or less conservative scale cuts; weak lensing, spectroscopic galaxy clustering and their galaxy-galaxy lensing combination, including cross-correlation.
We remark here that we are not intending to simply combine all datasets together, as this would not allow us to understand the impact that each dataset has on the constraints, mixing datasets which may have a different degree of reliability, due to different levels of systematics. Rather, we investigate different choices (for example different scale cuts when combining CMB data, which highlight whether Planck or ACT/SPT are dominating the constraints) and add, wherever possible, new data sets one by one. We will start with CMB data, and then adding background data; next, we will consider only late-time probes coming from galaxy surveys. Finally, we will consider both high redshift and low redshift probes.

\subsection{Cosmic Microwave Background}\label{sec:CMB}
We test our tomographic CDE model on \textit{Planck} 2018 low-$\ell$ and high-$\ell$ TT, TE and EE CMB spectra \cite{Planck:2018vyg}, covering the multipole range $2 \leq \ell \leq2508$ for the TT, and $2 \leq \ell \leq1996$ for the TE and EE power spectra. We call this dataset Planck, in short. For completeness, in one of our datasets, we also include the CMB lensing power spectrum in the multipole range $8 \leq \ell \leq 400$ and combine it with the Planck TT,TE,EE data. We denote it as Planck+PlanckLens. 

We also study four additional combinations of CMB data, by including the information from the ACT and SPT-3G surveys, which cover a larger multipole range than {\it Planck}: 

\begin{enumerate}
    \item Atacama Cosmology Telescope (ACT) Data Release 4 \cite{ACT:2020gnv}:\\
    \\We use CMB temperature anisotropy and polarisation (TT, EE and TE) power spectra from the latest ACT DR4, measured with the ACTPol receiver. We use the full $\ell$-range of $600 \leq \ell \leq 4125$ for the TT power spectra and $350 \leq \ell \leq 4125$ for the TE/EE power spectra. Data from ACT complements the Planck dataset in the high $\ell$ range. Following the recommendation of \cite{ACT:2020gnv}, we use a Gaussian prior on the optical depth of reionisation,  $\tau_\mathrm{reio} = 0.065 \pm 0.015$, which is only relevant when ACT is employed without including the low multipoles from {\it Planck} since in this case $\tau_\mathrm{reio}$ is basically unconstrained by the likelihood. All the results reported in this paper, though, are obtained by combining the ACT data with the low-multipole information from {\it Planck}, so the impact of this prior is in practice negligible. Along the paper and in all our figures and tables we denote the ACT dataset simply as ACT.
    \item \textit{Planck} + ACT (TT cut at $\ell_{\rm min}=1800$):\\
    \\We combine ACT and \textit{Planck} data. After a Fisher matrix analysis carried out by \cite{ACT:2020gnv}, the authors  found it necessary to truncate the ACT TT power spectrum at multipoles $\ell < 1800$ to avoid double counting issues caused by the overlap of the multipole ranges covered by the two experiments. The addition of the {\it Planck} information has the potential to greatly tighten cosmological constraints with respect to the case in which only ACT data is considered, see e.g. Figs. 12 and 17 in \cite{ACT:2020gnv}. We denote this truncated ACT dataset as ACT1800, and its combination with the full {\it Planck} temperature and polarisation likelihoods as Planck+ACT1800. 
		
    \item \textit{Planck} + ACT1800  + SPT-3G:\\
    \\As a next step, we include TE and EE power spectra from the South-Pole 3G telescope \cite{SPT-3G:2021eoc}, covering a multipole range of  $300 \leq \ell \leq 3000$. Constraints from SPT-3G CMB data have shown to be in good agreement with \textit{Planck}, hence providing further independent validation of \textit{Planck} results. In addition, SPT data has much better constraining power at high $\ell$'s compared to $\textit{Planck}$ and also covers a separate area of the sky than ACT, so is highly complementary to the former two datasets \cite{SPT-3G:2021wgf}. Since the overlap between SPT-3G and \textit{Planck} is small and there is no overlap of the SPT survey area with ACT, there is no need to truncate their power spectra when used in combination. We call this dataset Planck+ACT1800+SPT. 
		
    \item  \textit{Planck} (TT cut at $\ell_{\rm max}=650$) + ACT +SPT-3G:\\
    \\Following \cite{Hill:2021yec}, we also consider an alternative way to combine Planck and ACT: we combine again the ACT, \textit{Planck} and SPT-3G data, but this time limiting the \textit{Planck} TT power spectrum to $\ell_{max}= 650$. The exclusion of higher multipoles of the \textit{Planck} TT spectrum has shown to give rise to a preference for early dark energy models and a higher value of $H_0$, see \cite{Hill:2021yec,Poulin:2021bjr,Smith:2022hwi}. Hence, it will be interesting to look into whether this dataset suggests a stronger preference for CDE as well. We denote this dataset as Planck650+ACT+SPT. With respect to the previous choice 3, this combination includes a larger part of ACT data.
\end{enumerate}
\subsection{Baryonic Acoustic Oscillations}
We use the following BAO data points:
\begin{itemize}
    \item $D_V/r_s (z=0.122)$ \cite{Carter:2018vce}, the effective dilation scale obtained by  data from 6dFGS and SDSS Main Galaxy Sample.
    \item $D_M/r_s (z=0.32)$ and $r_sH(z=0.32)$ \cite{2017MNRAS.465.1757G} from the BOSS DR12 LOWZ samples.
    \item $D_M/r_s (z=0.57)$ and $r_sH(z=0.57)$ \cite{2017MNRAS.465.1757G} from the Baryonic Oscillation Spectroscopic Survey (BOSS) DR12 CMASS samples.
    \item $D_V/r_s (z=0.44,0.60,0.73)$ \cite{2012MNRAS.425..405B} from the WiggleZ Dark Energy Survey. 
    \item $D_M/r_s (z=0.835)$ \cite{DES:2021esc} from the Dark Energy Survey (DES) Year3 Data Release.
    \item $D_M/r_s (z=1.19,1.50,1.83)$ and $r_sH(z=1.19,1.50,1.83)$ \cite {Gil-Marin:2018cgo} from the extended BOSS (eBOSS) DR14 quasar sample (DR14Q).
    \item $D_M/r_s (z=2.34)$ \cite{deSainteAgathe:2019voe} obtained by the correlations between Ly-$\alpha$ absorption and eBOSS quasar spectra.

\end{itemize}
where $D_M(z)$ is the comoving angular diameter distance, $r_s$ the sound horizon at the baryon-drag epoch, and 
\begin{equation}
D_V(z)=\left[D_M^2(z)\frac{cz}{H(z)}\right]^{1/3}
\end{equation}
the so-called dilation scale. The covariance matrices of the BOSS DR12 LOWZ and CMASS samples and eBOSS DR14Q have been duly taken into account.
\subsection{Type Ia supernovae}
We employ the following SNe1a data:
\begin{itemize}
    \item Six effective points on the Hubble rate, i.e. $E(z)=H(z)/H_0$, and their associated covariance. They compress the data from the 1048 SNe1a of the Pantheon compilation \cite{Scolnic:2017caz} and the 15 SNe1a at $z > 1$ from the Hubble Space Telescope Multi-Cycle Treasury programs \cite{Riess:2017lxs}. The compression effectiveness of the information contained in such SNe1a samples is excellent, see Ref. \cite{Riess:2017lxs}.
		
    \item Combination of the lightcurve data from 251 spectroscopically confirmed SNe1a in the redshift range of $0.02 < z < 0.85$, measured by the Dark Energy Survey Supernova Programme (DES-SN) \cite{2019ApJ...872L..30A}.
\end{itemize}
\subsection{Cosmic Chronometers}
We use the 31 data points of the Hubble function $H(z)$ in a redshift range of $0.07 \leq z \leq 1.905$, obtained by making use of passively-evolving galaxies and the differential age technique \cite{Jimenez:2001gg}. These values were calculated without assuming any fiducial cosmology. We use the data on cosmic chronometers (CCH) compiled in Table 1 of \cite{Gomez-Valent:2018gvm}, see references therein. 
\subsection{Redshift-Space Distortions}
We use the following data from redshift-space distortions (RSD), expressed in terms of the product of the growth rate $f(z)\equiv d\ln\delta_m/d\ln a$ \footnote{We note that in the case of a coupled quintessence model, the modified evolution of the density contrast $\delta_\mathrm{cdm}$ adds a contribution to $f$, such that the effective growth rate becomes $f_\mathrm{eff}=f-\kappa\phi\prime\beta/\mathcal{H}$ \cite{Martinelli:2019dau}. However, we have explicitly checked that this correction is very small: using a test value of $\beta=0.04$, the ratio $|f_\mathrm{eff} - f|/f  \sim 10^{-3}$ in the redshift range of the RSD data points, while the uncertainties of the data are on the order $\sigma_{f\sigma_8}/f\sigma_8 \sim 0.1-0.2$.} and the {\it rms} of mass fluctuations at scales of $R_8=8h^{-1}$ Mpc, $\sigma_8$:
 \begin{itemize}
        \item $f\sigma_8 (z=0.03)$ \cite{Qin:2019axr} derived by combining results from the 2MTF and 6dFGS surveys.
        \item $f\sigma_8 (z=0.1)$ \cite{2018ApJ...861..137S} from the Sloan Digital Sky Survey (SDSS) Data Release 7.
        \item $f\sigma_8 (z=0.18,0.38)$ \cite{Simpson:2015yfa} \cite{2013MNRAS.436.3089B} from the Galaxy and Mass Assembly survey (GAMA).
        \item $f\sigma_8 (z=0.22,0.41,0.60,0.78)$ \cite{2012MNRAS.425..405B} from WiggleZ.
        \item $f\sigma_8 (z=0.32)$ \cite{2017MNRAS.465.1757G} from BOSS DR12 LOWZ.
        \item $f\sigma_8 (z=0.57)$ \cite{2017MNRAS.465.1757G} from BOSS DR12 CMASS.
        \item $f\sigma_8 (z=0.60,0.86)$ \cite{2018A&A...619A..17M} from VIMOS Public Extragalactic Redshift Survey.
        \item $f\sigma_8 (z=0.77)$ \cite{Guzzo:2008ac} from VIMOS VLT Deep Survey.
        \item $f\sigma_8 (z=1.19,1.50,1.83)$ \cite {Gil-Marin:2018cgo} from eBOSS DR14Q.
        \item $f\sigma_8 (z=1.36)$ \cite{Okumura:2015lvp} from Subaru FMOS galaxy redshift survey (FastSound).
    \end{itemize}

We correct for the Alcock-Paczynski effect by multiplying $f\sigma_8$ by the ratio $[H(z)D_A(z)]/[H^{\rm fid}(z)D^{\rm fid}_A(z)]$, as in \cite{Macaulay:2013swa}, where {\it fid} denotes the fiducial cosmology used in the survey. In the CMASS, LOWZ and eBOSS DR14Q samples we take into account the correlations between the BAO and RSD data through the corresponding covariance matrices.
\subsection{SH0ES}
We use the most updated prior on the Hubble constant obtained by the SH0ES team using the cosmic distance ladder method, $H_0=(73.04\pm 1.04)$ km s $^{-1}$Mpc$^{-1}$ \cite{Riess:2021jrx}, which is in ${\sim} 5\sigma$ tension with the value obtained by the $\textit{Planck}$ collaboration under the assumption of the flat $\Lambda$CDM model.
\subsection{Weak Lensing}\label{sec:WLdata}
We employ cosmic shear data from the latest data release of the Kilo-Degree Survey, KiDS-1000 \cite{Kuijken:2019gsa}, covering an effective area of 777deg$^2$ with photometric galaxies spanning the redshift range $0.1 < z \leq 1.2$. We have validated our pipeline by conducting runs with the $\Lambda$CDM model, making use of our modified version of \texttt{CLASS} (while setting $\beta(z)=0$) and checking that we are able to reproduce the results that are reported in \cite{2021A&A...646A.140H}. In this study, we follow the fiducial analysis settings keeping an $\ell$ range of $\ell\in[100,1500]$ in the calculation of the cosmic shear angular power spectrum $C^\ell_{\epsilon\epsilon}$. Here, we acknowledge that although fitting functions exist to model the nonlinear matter power spectrum  $P_{nl}(k)$ for a constant coupled dark energy model \cite{Casas:2015qpa}, an accurate recipe for the tomographic coupling case has yet to be developed and would need Nbody simulations for varying coupling which are not publicly available at present; we leave this task for future work. For this reason, we conservatively use only scales up to $\ell < 1500$; while this already cuts a large part of non-linear scales that may be interesting to include in future studies, we also use $\textsc{HMCode}$ \cite{Mead:2016zqy} to approximate $P_{nl}(k)$ within this range; this approximation is good enough for our purposes, under the assumption that in this regime and for the values of $\beta$ considered in this paper, the difference between the non-linear spectra obtained with $\textsc{HMCode}$ corrections, and the spectra in a CDE model, is small. We leave a more refined analysis at non-linear scales, including smaller scales than the ones included here, for future studies. 
\subsection{Spectroscopic Galaxy Clustering}
We consider spectroscopic galaxy clustering data from the SDSS-III Baryonic Oscillation Spectroscopic Survey \cite{BOSS:2016wmc} DR12, following the methodology of \cite{2017MNRAS.464.1640S}, where they calculate the anisotropic redshift space correlation function $\xi_{gg}(s,\mu,z)$ in 3 three-dimensional `wedges' (transverse, intermediate and parallel to the line of sight), split into 2 tomographic bins of $0.2 < z \leq 0.5$ and $0.5 < z \leq 0.75$.
\subsection{3x2 pt}
Subsequently, we use, for the first time, the cross-correlations of the aforementioned KiDS-1000 `source' galaxies, split into 5 tomographic bins,  with `lens' galaxies obtained by combining both BOSS DR12 and 2dFLenS \cite{Blake:2016fcm} data to obtain the galaxy-galaxy lensing angular power spectra $C_{n\epsilon}^\ell$ to constrain our model. The combination of cosmic shear, galaxy clustering and galaxy-galaxy lensing will then make up our 3x2pt probe. Since the overlap between the KiDS and BOSS surveys only accounts for 4\% of the BOSS footprint, the galaxy clustering covariance matrix has been assumed to be independent of the covariance matrix of cosmic shear and galaxy-galaxy lensing, which was derived analytically in \cite{Joachimi:2020abi}.  In the case of galaxy-galaxy lensing, we follow \cite{2021A&A...646A.140H} to limit the contribution of nonlinear scales with $k \gtrsim 0.3$ $h$Mpc$^{-1}$. The authors argue that beyond these scales, their theoretical model of the galaxy-galaxy lensing power spectrum becomes significantly inaccurate due to nonlinearities. This cut in scale corresponds, based on calculations of the data and the realspace correlations functions $\xi_\pm (s,\mu)$, to an $\ell$ range of $\ell\in[100,300]$. We will thus also use this $\ell$ range in our 3x2pt analysis.

\section{Methodology}\label{sec:Method}
We use a modified version of \texttt{CLASS} with the model described in Sec. \ref{sec:CDE}. Our code allows us to solve the Einstein-Boltzmann system of cosmological equations, to obtain the theoretical quantities of interest at the background and linear perturbation levels, which is crucial to confront the model with observations and subsequently constrain model parameters. The code has been validated against the one used in \cite{1207.3293,Pettorino:2013oxa,Planck:2015bue}, for constant coupling (and previously with a modified version of CMBfast used in \cite{1111.1404}). The present version allows us to reach exactly $\Lambda$CDM in the limit of constant potential and zero coupling.

We employ \texttt{MontePython} \cite{Audren:2012wb,Brinckmann:2018cvx} to explore the parameter space of our CDE models. We use the regular Metropolis-Hastings sampling algorithm \cite{1953JChPh..21.1087M,Hastings:1970aa}, and stop the Monte Carlo when the Gelman-Rubin convergence statistic R$-1 < 0.02$ \cite{R2:1992,R1:1997}. On top of the $\beta_i$'s, we vary the reduced baryon density, $\omega_\mathrm{b}$, the reionization depth, $\tau_{\rm reio}$, the cosmological constant $V_0$, the initial CDM density at $z_{\rm ini}=10^{14}$, and the parameters that characterise the shape of the primordial power spectrum, i.e. its amplitude $A_\mathrm{s}$ and the spectral tilt $n_\mathrm{s}$. Notice that we vary directly the initial CDM density instead of the reduced CDM density, $\omega_{\rm cdm}$. This allows us to avoid the use of the shooting method and speed up the Monte Carlo runs, as in \cite{Gomez-Valent:2020mqn,Gomez-Valent:2022bku}. In this paper, we are interested in setting the initial conditions in the past, evolving quantities under the effect of the varying coupling, in order to see the effect on present quantities. We keep the Hubble parameter, $H_0$, $\omega_{\rm cdm}$, the root mean square of mass fluctuations at scales of $8h^{-1}$ Mpc, $\sigma_8$, and $S_8$ as derived parameters. We consider two massless neutrinos and one massive neutrino of 0.06 eV. For the initial conditions of the scalar field, we are allowed to set $\phi_{\rm ini}=0$ and $\phi_{\rm ini}^\prime=0$, since the scalar field has no dynamics during the radiation-dominated epoch due to the null impact of the potential in that era, and the equations do not depend on $\phi$, but only on its time derivatives.

We choose uninformative flat priors for all the parameters, wide enough to not influence the posterior distribution. In particular, we vary the couplings in the range $\beta_i\in[0,\beta_{\rm max}]$, where the value of $\beta_{\rm max}$ is chosen depending on the particular data set. It has to be sufficiently large to not cut the tail of the posterior artificially. The Monte Carlo Markov chains are analysed using the $\texttt{GetDist}$ \cite{Lewis:2019xzd} Python package.

\subsection{Binning}
When binning the coupling, we need to make a choice on the bins to be considered. We start from the consideration that, depending on the datasets we include in the analysis, different choices may be more or less adequate. For example, if we are interested in including CMB data, it is interesting to consider bins which differentiate between late and early times. If instead we only include late-time probes, we may rather be interested in increasing the number of bins within a redshift range which overlaps with where the data are available. While there is no unique choice, we here test different typical choices, depending on the datasets.
Here we list the different binning models we test, and the datasets we have chosen to constrain each of them.


\subsubsection{3-bin tomographic $\beta$}
We first consider CMB data alone, to evaluate the impact of different choices which can be made when using Planck alone, or in combination with ACT and SPT. For this purpose, we test a binning with 3 tomographic bins, using bin edges $\{0,100,1000\}$. This choice roughly identifies the structure formation, post-recombination and pre-recombination eras. We fix $s_1=s_2=0.03$ to allow for a smooth, gradual transition between amplitudes in the various bins. Here we study if the data prefer a different value of the coupling in the pre- and post-recombination epochs, allowing also for a change in $\beta$ at $z<100$. 

\begin{figure*}[t!]
    \centering
    \includegraphics[width=\linewidth]{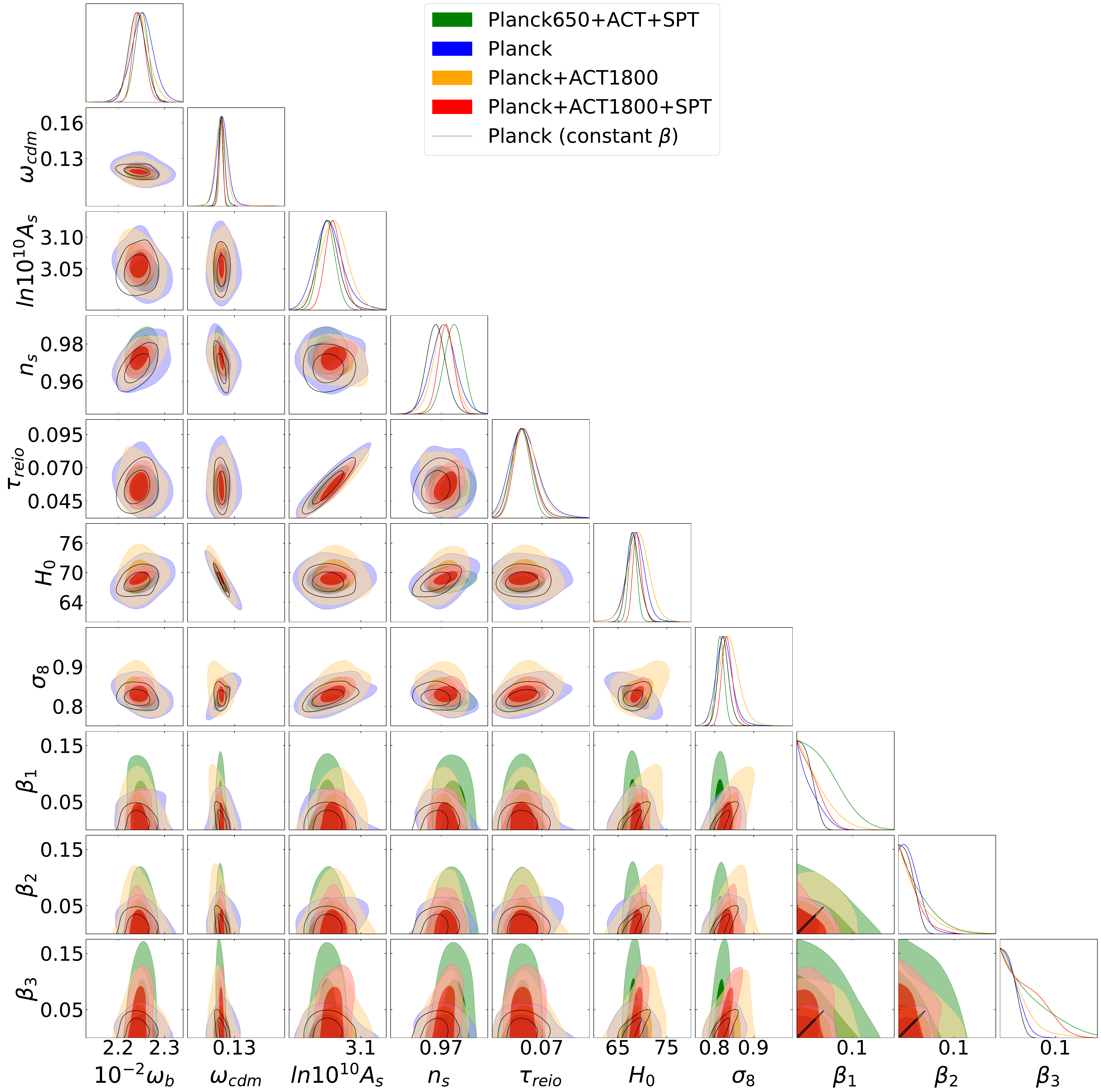}
    \caption{Triangular plot of 68\%  and 95\% C.L. posterior distributions obtained with the CMB datasets Planck650+ACT+SPT (green), Planck (blue),  Planck+ACT1800 (yellow) and Planck+ACT1800+SPT (red) for the 3-bin tomographic model. The binning is defined by the edges $z=\{0,100,1000\}$. We include the contours derived for a constant $\beta$ model, using Planck data, in black. In this case, the contours for $\beta_{1-3}$ correspond to the same constant $\beta$.}
    \label{fig:full_cmb_contour}
\end{figure*}

We test this model with CMB data alone, using the datasets described in Sec.~\ref{sec:CMB}, namely:  Planck, Planck+PlanckLens, Planck650+ACT+SPT, Planck+ACT1800 and Planck+ACT1800+SPT, since we would like to test the constraining power of these alternative CMB data sets as compared to a purely Planck dataset.  
\subsubsection{7-bin tomographic $\beta$}\label{sec:method7}
Next, we aim at combining CMB data with background datasets from BAO, supernovae and cosmic chronometers, with or without RSD. We then test a binning with 7 tomographic bins, using bin edges $\{0,1,2,5,100,500,1000\}$. In this case we use $s_i=10$ for $i\in[1,3]$, and $s_i=0.03$ for $i\in[4,6]$. This finer binning will let us study to what extent the constraints on the coupling are sensitive to the rich variety of data at low redshifts and to the physical processes around the decoupling time of the CMB photons. Our definition of the tomographic bin edges is largely motivated by the redshift ranges that are being covered by the datasets we employ. 

We begin with a baseline CMB dataset, to which we include other probes to see how constraints might be affected, namely: BAO+SNe1a+CCH datasets (denoted as `BSC'), RSD data, and finally a SH0ES prior to see if its inclusion could potentially increase the coupling strength at various epochs.

Our two baseline datasets are Planck+ACT1800+SPT and Planck650+ACT+SPT. It is interesting to run our analyses on both combinations of CMB data cuts (cut on ACT versus cut on Planck) to study how these different cuts impact the constraints on our model. 

We list the datasets here, and in brackets the shorthand notation we employ in the following sections: 
\begin{enumerate}
    \item Planck+ACT1800+SPT+BSC
    \item  Planck+ACT1800+SPT+BSC+RSD
    \item  Planck+ACT1800+SPT+BSC+SH0ES
    \item  Planck+ACT1800+SPT+BSC+RSD+SH0ES
    \item Planck650+ACT+SPT+BSC
    \item Planck650+ACT+SPT+BSC+RSD
    \item  Planck650+ACT+SPT+BSC+SH0ES
\end{enumerate}
\subsubsection{4-bin tomographic $\beta$}
Lastly, we concentrate on late-time probes only; for this purpose, we test a binning with 4 tomographic bins, using bin edges $z = \{0,0.5,1,2\}$ and $s=10$ in the case of weak lensing (KiDS-1000 cosmic shear), BOSS spectroscopic galaxy clustering and the 3x2pt datasets. Since the redshift ranges of KiDS-1000 source galaxies are between $0.1 \leq z \leq 1.2$ and the BOSS and 2dFLenS lens galaxies are between $0.2 < z < 0.75$, we alter our tomographic binning model to only vary $\beta(z)$ at this range, ensuring that the data can be fully exploited to constrain our model. Moreover, since the first three tomographic bins are very narrow in this binning model and only vary at low redshifts, the impact of larger $\beta_i$'s on the matter power spectrum is less significant. 

 \begin{figure}[t!]
    \centering
    \includegraphics[width=\linewidth]{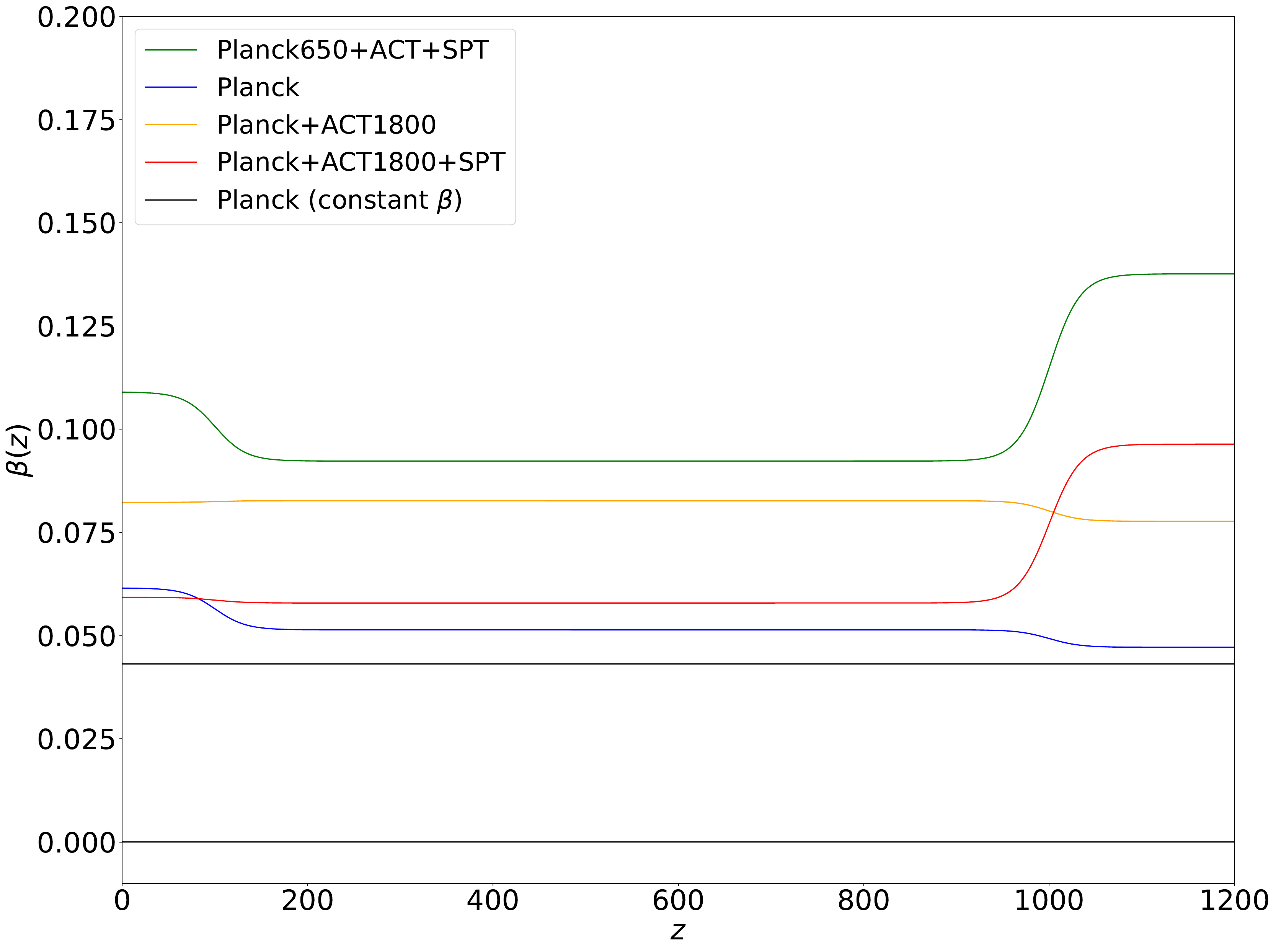}
    \caption{95\% C.L. on $\beta(z)$ for the 3-bin tomographic model. The binning is defined by edges $z=\{0,100,1000\}$. For reference, we also include the constraint for the case of a constant coupling $\beta$, obtained with Planck data, in black.}
    \label{fig:beta_limits_cmb}
\end{figure}

We use the cosmic shear, spectroscopic galaxy clustering, and 3x2pt datasets to test both a constant and tomographic coupling model, and subsequently combine cosmic shear with the Planck+ACT1800+SPT+BSC dataset in a tomographic framework, to see if $\beta(z)$ can become more constrained at higher redshifts. We only do this for cosmic shear and not galaxy clustering and 3x2pt, to avoid double counting when combining galaxy clustering with BAO and RSD. Additionally, when using only cosmic shear, we follow \cite{2021A&A...646A.140H} by imposing a flat prior on $H_0$ encompassing $\pm5\sigma$ of distance ladder constraints \cite{Riess:2016jrr}, given by $ 64  < H_0 < 82$ km/s/Mpc. 

\begin{figure*}[t!]
    \centering
    \includegraphics[width=\linewidth]{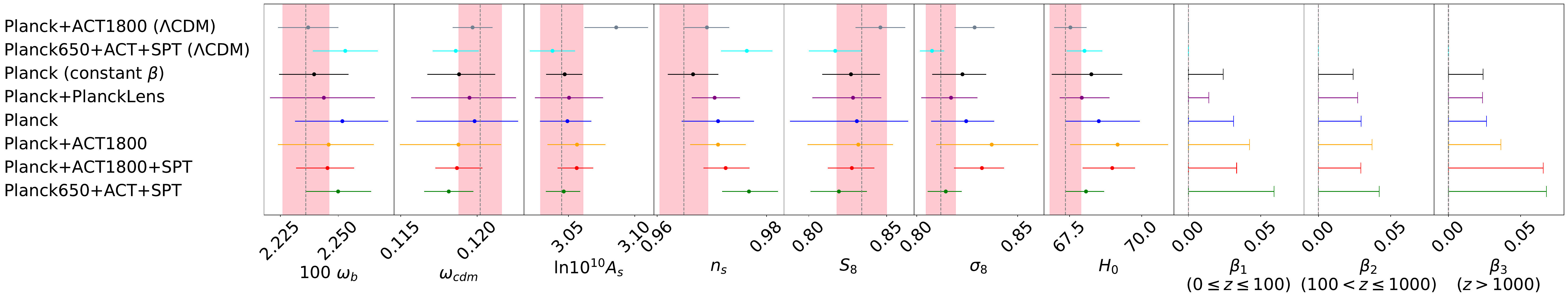}
    \caption{Comparison of the mean and 68\% C.L. of the various cosmological parameters, for the different CMB datasets with a 3-bin model. Here the grey vertical dashed lines and pink bands denote respectively the mean and 1$\sigma$ of $\textit{Planck}$ $\Lambda$CDM TTTEEE+lowE fiducial cosmology (cf. Table 2 of \cite{Planck:2018vyg}), and the results obtained for the Planck+ACT1800 ($\Lambda$CDM) dataset in grey were taken from Table 4 of \cite{ACT:2020gnv}. We also make a comparison to the constant $\beta$ model with Planck data (in black).}
    \label{fig:horiz_bestfit_cmb}
\end{figure*}

Weak lensing has been used to constrain a constant CDE model in previous literature \cite{Planck:2015bue}, however it is the first time we cross-correlate it with full-shape galaxy clustering to probe coupled dark energy models, hence it is interesting to probe their combined constraining power. 
\section{Results} \label{sec:Results}
\subsection{CMB with 3-bin Tomographic $\beta$}
We show in Fig.~\ref{fig:full_cmb_contour} the posterior distribution of the standard cosmological parameters $\mathbf{\theta} = \{ \omega_\mathrm{b}, \omega_{\rm cdm}, \ln{10^{10}A_\mathrm{s}}, n_\mathrm{s},\tau_\mathrm{reio}, H_0, \sigma_8 \} $, as well as the constraints on $\beta_{1-3}$, for a 3-bin tomographic model. In this model, we only consider CMB datasets and we fit our model with Planck, Planck+ACT1800, Planck+ACT1800+SPT and Planck650+ACT+SPT (cf. Sec. \ref{sec:CMB}). For a detailed list of the best-fit, mean and 68\% C.L. uncertainties for each parameter, we refer the reader to Table~\ref{table:cmb_bestfit} in Appendix \ref{sec:AppendixA}.  For the results and analysis obtained with a Planck+PlanckLens dataset, we refer the reader to Appendix \ref{sec:AppendixB1}.

We see that the differences between the contour plots obtained from the analyses with Planck alone and Planck+ACT1800 when the scale cut includes more of Planck and less of ACT data are not substantial, and also do not differ considerably from the one derived with a constant coupling model. This is expected since, as already mentioned, \textit{Planck} data carries the bulk of the statistical power. When instead high multipoles of the Planck spectrum are excluded, i.e. when we consider the Planck650+ACT+SPT dataset, the constraints loosen considerably for every coupling coefficient $\beta_{1-3}$. 
In all cases, we find that the interaction strength is compatible with 0 (i.e. with $\Lambda$CDM) at 68\% C.L.

There is a positive correlation between $H_0$, $\sigma_8$ and the coupling coefficients $\beta_{1-3}$, especially in the case Planck+ACT1800. The correlation is stronger for $\beta_1$, i.e., in the redshift range $0 < z < 100$. This behaviour ties in with the results presented in Fig.~\ref{fig:3binSpectra}: an increase in $\beta_1$ leads to an increase of the matter power spectrum and hence, to $\sigma_8$, and requires larger values of $H_0$ to respect the position of the first acoustic peak of the CMB temperature spectrum. 

\begin{figure*}[t!]
    \centering
    \includegraphics[width=\linewidth]{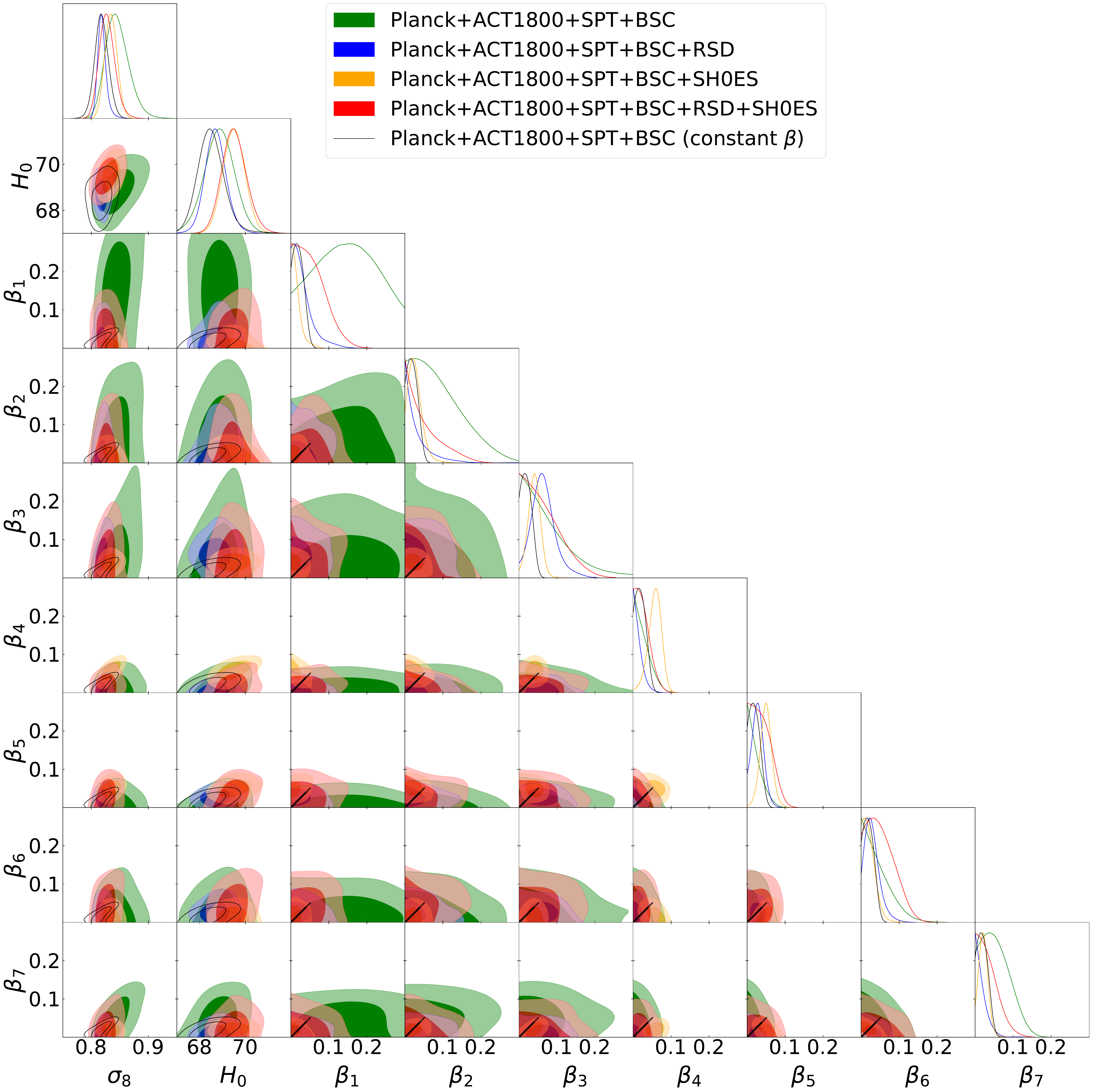}
    \caption{Triangular plot of 68\% and 95\% C.L. posterior distributions of $\sigma_8$, $H_0$, and the 7 tomographic coupling coefficients $\beta_{1-7}$ derived from datasets Planck+ACT1800+SPT+BSC (green), Planck+ACT1800+SPT+BSC+RSD (blue), Planck+ACT1800+SPT+BSC+SH0ES (yellow) and Planck+ACT1800+SPT+BSC+RSD+SH0ES (red). For reference, the binning is defined by edges $z=\{0,1,2,5,100,500,1000\}$. We include, in black lines, the contours obtained for a constant $\beta$ case with Planck+ACT1800+SPT+BSC data. In this case, the contours for all $\beta_{1-7}$ are the same.}
    \label{fig:actpolcut_full}
\end{figure*}

\begin{figure}[t!]
    \centering
    \includegraphics[width=\linewidth]{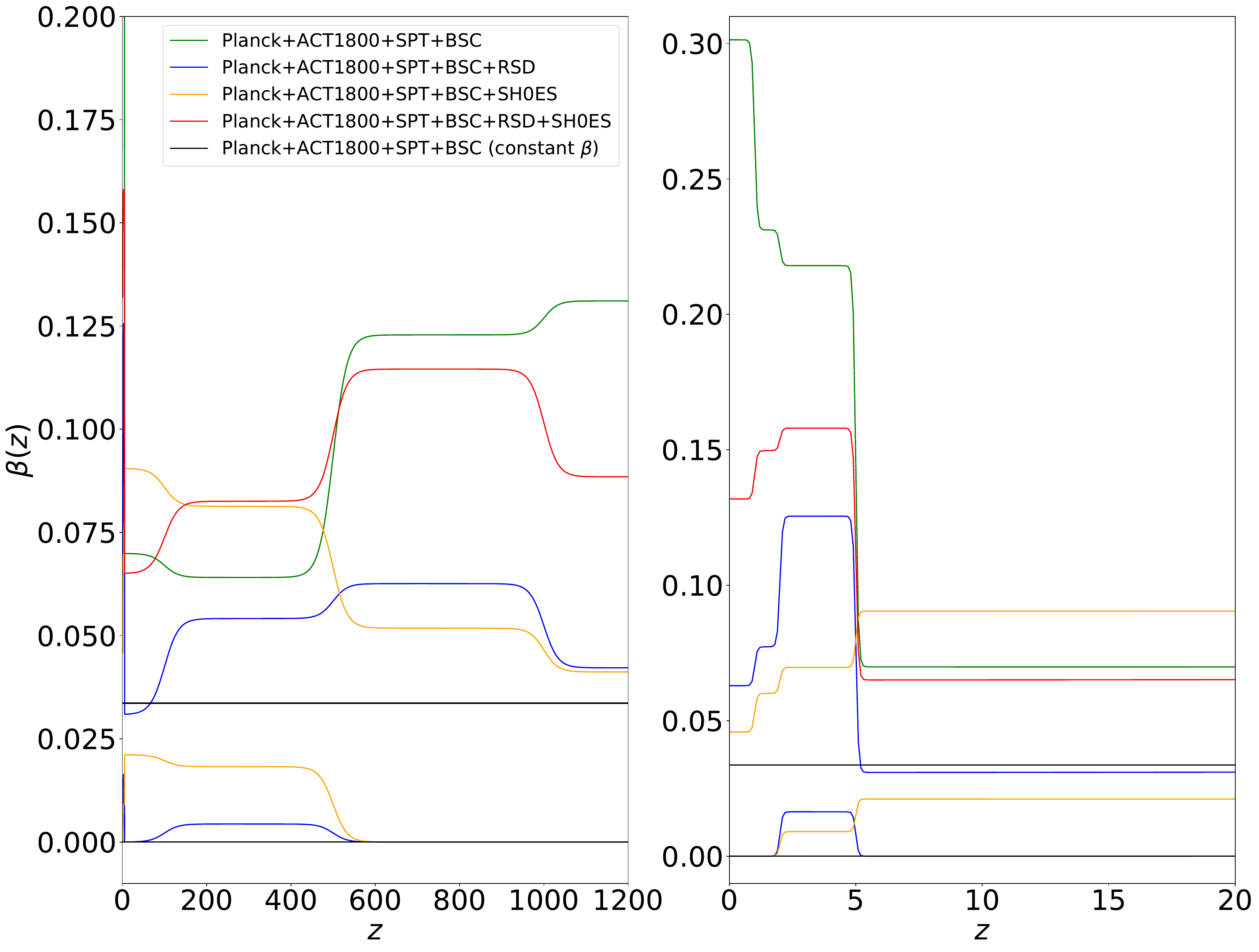}
    \caption{95\% C.L. on $\beta(z)$ for a 7-bin coupling model constrained with different datasets that incorporate the baseline combination Planck+ACT1800+SPT. The right plot is a low-redshift zoom of the plot on the left. For reference, we include the constraints for a constant $\beta$ with Planck+ACT1800+SPT+BSC, in black. The binning is defined by edges $z=\{0,1,2,5,100,500,1000\}$. }
    \label{fig:actpolcut_beta_lims}
\end{figure}

In Fig.~\ref{fig:beta_limits_cmb} we plot the 95\% C.L. constraints on the evolution of the coupling strength $\beta(z)$, reconstructed from $\beta_{1-3}$ via \eqref{eq:binbeta}: constraints are derived for the various CMB datasets, and compared to the case of a constant coupling, as studied in \cite{Gomez-Valent:2020mqn,Gomez-Valent:2022bku}. We see that in all cases, CMB constraints on $\beta(z)$ are weaker than for a constant coupling, going from $\beta_3$ < 0.045 at 95\% C.L. to a maximum of $\beta_3$ < 0.14 at 95\% C.L. for the case of Planck650+ACT+SPT. Interestingly, adding ACT and SPT data to Planck loosens constraints at all epochs, and for all scale cuts, with respect to Planck alone. Constraints in $\beta_3$ bin, around recombination mainly come from Planck data. This is due to the fact that Planck data tightly constrain the amplitude of the first acoustic peak, to which, as we have seen in Fig.~\ref{fig:3binSpectra}, a change in $\beta_3$ is most sensitive. 
The inclusion of SPT further loosens constraints around recombination.

Fig.~\ref{fig:horiz_bestfit_cmb} summarises our results for the 3-bin scenario: we show the mean and 68\% C.L. of the cosmological parameters of our 3-bin model, as constrained by different choices of CMB datasets, and compare them with those obtained for $\Lambda$CDM and for a constant coupling ($\beta_1=\beta_2=\beta_3=\beta$) model. The pink band represents the mean and 1$\sigma$ uncertainty on $\Lambda$CDM cosmological parameters as obtained from \textit{Planck}. We note that ACT and SPT are in better agreement with Planck data within a CDE cosmology (red and orange vs blue lines) than in $\Lambda$CDM (grey and cyan vs pink band). As shown before, when the coupling is allowed to vary in redshift, as in a tomographic model, it is less constrained than when it is forced to be constant at all redshift bins, and in particular near recombination. The inclusion of ACT and SPT further loosens the constraints.

Interestingly, we see that a tomographic CDE model shifts $H_0$ towards slightly higher mean values for all datasets, when compared to $\Lambda$CDM. Using the Planck dataset as a basis for comparison, we find that a tomographic model gives very similar results compared to the constant coupling case, with  $H_0=(68.15_{-1.43}^{+1.21})$ km/s/Mpc for the constant coupling model and $H_0=(68.51_{-1.17}^{+1.42})$ km/s/Mpc for the tomographic case. Including ACT, as in Planck+ACT1800, increases this value to $H_0=(69.17^{+1.75}_{-1.65})$ km/s/Mpc. If we take the results obtained only with Planck and compare them to the SH0ES value, the Hubble tension decreases from ${\sim} 4.8\sigma$ (assuming $\Lambda$CDM) to ${\sim} 2.9\sigma$ (assuming CDE with $\beta=$ const.). In the case of the  tomographic model, the $H_0$ tension decreases even more, to ${\sim} 2.7\sigma$. However it is worth noting that the uncertainties in $H_0$ is increased by a factor of 2 going from $\Lambda$CDM to a CDE model, which is the main cause of this reduction of the tension.

Finally, we also note that the tension between the value of $n_\mathrm{s}$ measured with Planck and Planck650+ACT+SPT when assuming $\Lambda$CDM disappears in the context of a tomographic CDE model; the error bars overlap (dark blue and green) and thus the corresponding values of $n_s$ are compatible. Similarly, the mild tension between the value of $A_s$ found with Planck and Planck+ACT1800 in the $\Lambda$CDM is also washed out in the tomographic model (cf. the dark blue and yellow bars).

In summary, the results obtained with the 3-bin tomographic model when using only CMB data allow for larger couplings at all epochs; none of the current datasets detects a non-zero coupling in any bin, and data are consistent with $\Lambda$CDM predictions; the CDE model though, allows for slightly larger values of $\sigma_8$ and $H_0$ with respect to $\Lambda$CDM. On the other hand, a non-zero coupling works to decrease $\omega_{\rm cdm}$, which compensates the increase in $\sigma_8$ to give a value of $S_8$ similar to the one we would have for $\textit{Planck}$ $\Lambda$CDM. We will see in the next section whether this is still compatible with other background datasets.

\subsection{7-bin Tomographic $\beta$}
It is natural to wonder whether shifts observed in $\sigma_8$ and $H_0$ and the corresponding slight decrease in $\omega_{\rm cdm}$ for a CDE model, are still compatible with background data, such as BAO and low redshift data from RSD. We present results for datasets 1-7 as listed in Sec.~\ref{sec:method7}, adapting the binning to a 7-bin model, which is finer at low redshifts. 
 We then add to the Planck+ACT1800+SPT baseline lower redshift data, namely BAO, SNe1a, cosmic chronometers, RSD and a SH0ES prior, and check their impact on the coupling strength.  

In Fig.~\ref{fig:actpolcut_full} we use Planck+ACT1800+SPT+BSC as a baseline and plot the constraints on $H_0$, $\sigma_8$ and the 7 coupling coefficients obtained for datasets 1-4, namely combining Planck$+$ACT1800$+$SPT with BSC, BSC$+$RSD, BSC$+$SH0ES and BSC$+$RSD$+$SH0ES. We recall that BSC refers to the combination BAO+SNe1a+CCH. A list of the best-fit, mean and 68\% C.L. uncertainties for each parameter is given in Table~\ref{table:actpolcut_bestfit} of Appendix \ref{sec:AppendixA}.

We see that when we combine our baseline CMB dataset with BSC data, the coupling strengths at low redshifts ($z<5$) are not well constrained (green contours for $\beta_1$, $\beta_2$ and $\beta_3$). The addition of either RSD or SH0ES leads to tighter constraints: $\beta_1 < 0.017$ for Planck+ACT1800+SPT+BSC+RSD+SH0ES with respect to $\beta_1 = 0.146^{+0.075}_{-0.081}$ for Planck+ACT1800+SPT+BSC at 68\% C.L. Couplings at larger redshifts are instead mainly unaffected with respect to Planck+ACT1800+SPT+BSC. We also see a considerable loosening of constraints for all parameters in general in the tomographic coupling case as compared to a constant coupling (in black), and a shift towards larger values of $H_0$ and $\sigma_8$. 

In Fig.~\ref{fig:actpolcut_beta_lims} we plot the 95\% C.L. on $\beta(z)$ as derived from the seven $\beta_i$ described above, for the baseline Planck+ACT1800+SPT with different background data, as compared to the constant coupling case with a Planck+ACT1800+SPT+BSC dataset. We see once again that the constraints on the coupling strength are weaker in the tomographic scenario, especially in high redshift bins, reaching $\beta_7 = 0.063^{+0.023}_{-0.054}$ at 68\% C.L. around recombination for Planck+ACT1800+SPT+BSC, with respect to $\beta_7 = 0.015^{+0.005}_{-0.014}$ for the same dataset, when a constant coupling is assumed (i.e. nearly 5 times greater). Interestingly, when either RSD or SH0ES data are considered on top of Planck+ACT1800+SPT (blue and yellow curves in Fig.~\ref{fig:actpolcut_beta_lims}, respectively), a non-null-coupling seems to be preferred in some redshift bins: in particular, the inclusion of a SH0ES prior appears to favour a non-zero coupling at more than 95\% C.L. for $5 < z < 500$ (right panel, zoom in z). This is consistent with the results found in \cite{Gomez-Valent:2020mqn}, who find a $>2\sigma$-level preference for a non-zero coupling coefficient when including a SH0ES prior and data from strong-lensing time delays from H0LICOW. However, when we use the combination RSD+SH0ES (in red) this evidence for a non-null coupling disappears.

For the results and analysis obtained with a CMB baseline Planck650+ACT+SPT cut, we refer the reader to Appendix \ref{sec:AppendixB2}.

\begin{figure*}[ht]
\begin{center}
    \begin{subfigure}[b]{\linewidth}
            \centering
            \includegraphics[width=\linewidth]{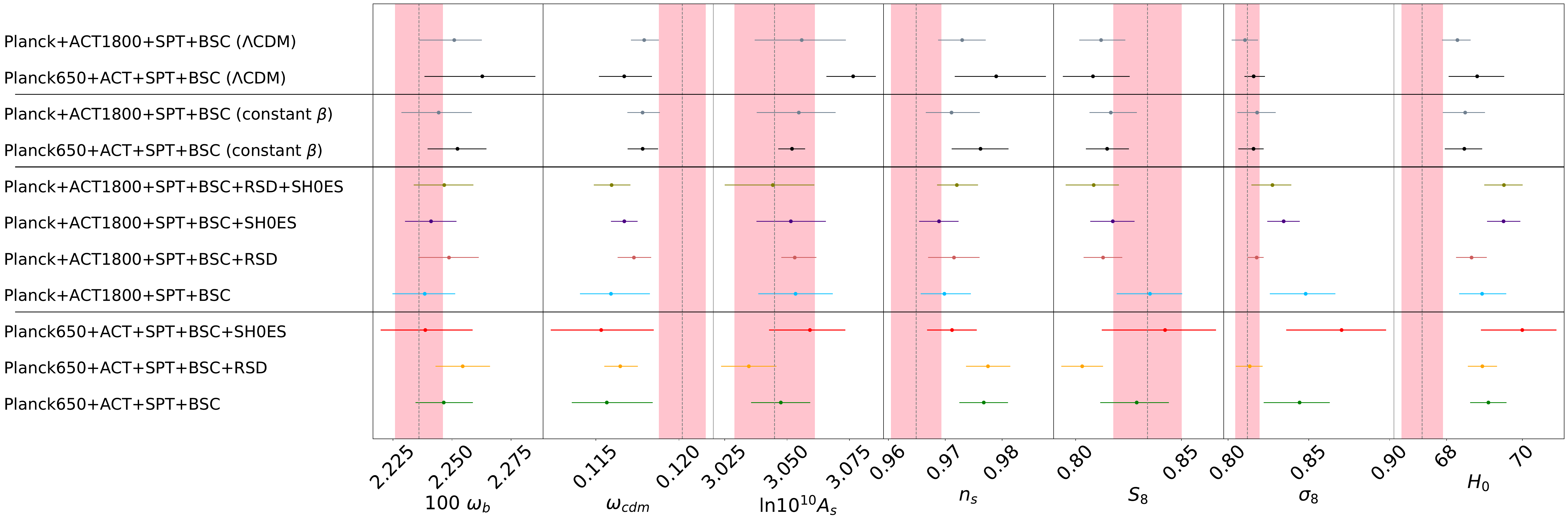}
        \end{subfigure}
        \vfill
        \vfill
        \begin{subfigure}[b]{\linewidth}  
            \centering 
            \includegraphics[width=\linewidth]{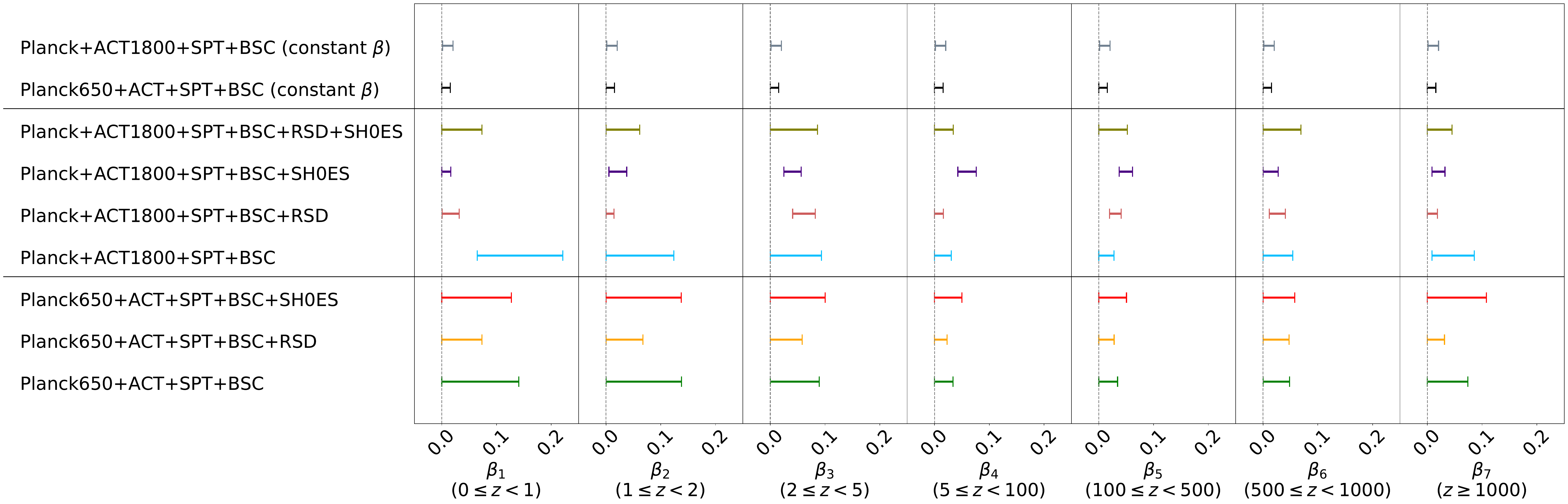}
        \end{subfigure}
        \caption
        {\small Comparison of the mean and 68\% C.L. values of the various cosmological parameters (top) and $\beta$ coupling coefficients (bottom), for the various datasets in a 7-bin tomographic CDE model. In the top plot, the grey vertical dotted lines and pink band denote respectively the mean and 1$\sigma$ of $\textit{Planck}$ fiducial cosmology. We also make a comparison with $\Lambda$CDM and a constant $\beta$ model, for datasets Planck+ACT1800+SPT+BSC (grey) and Planck650+ACT+SPT+BSC (black). } \label{fig:horiz_bestfit_planckcut}
\end{center}
\end{figure*}

In Fig.~\ref{fig:horiz_bestfit_planckcut} we present a comparison of the mean and 68\% C.L. of the various parameters derived for all the different datasets used to probe our 7-bin tomographic coupling model, comparing it to \textit{Planck} fiducial cosmology, a $\Lambda$CDM model with the same datasets, and a constant coupling model. Comparing across models, we see that a tomographic coupling generally gives lower values of $\omega_{\rm cdm}$, and higher values of $\sigma_8$ and $H_0$ than the constant coupling case by allowing for a stronger coupling at low and high redshifts. We see this reflected in our results for the case of $\sigma_8$: when we compare the same dataset analysed with a $\Lambda$CDM model and a CDE model, we see that with CDE, $\sigma_8$ (and thus $S_8$) increases, even if $\omega_\mathrm{cdm}$ decreases. For example, with a Planck+ACT1800+SPT+BSC dataset, $S_8=0.812 _{-0.010}^{+0.011}$ in $\Lambda$CDM, while $S_8=0.834^{+0.015}_{-0.017}$ in a tomographic CDE case.

The 7-bin model allows us to better understand how the coupling affects the universe dynamics at different redshifts. A larger coupling is allowed at low redshifts (z < 5) and high redshifts (z > 1000), outside of the period of large-scale structure formation in the matter-dominated epoch. 

\subsection{CDE with Weak Lensing and Galaxy Clustering}
\begin{figure*}[t!]
    \centering
    \includegraphics[width=\linewidth]{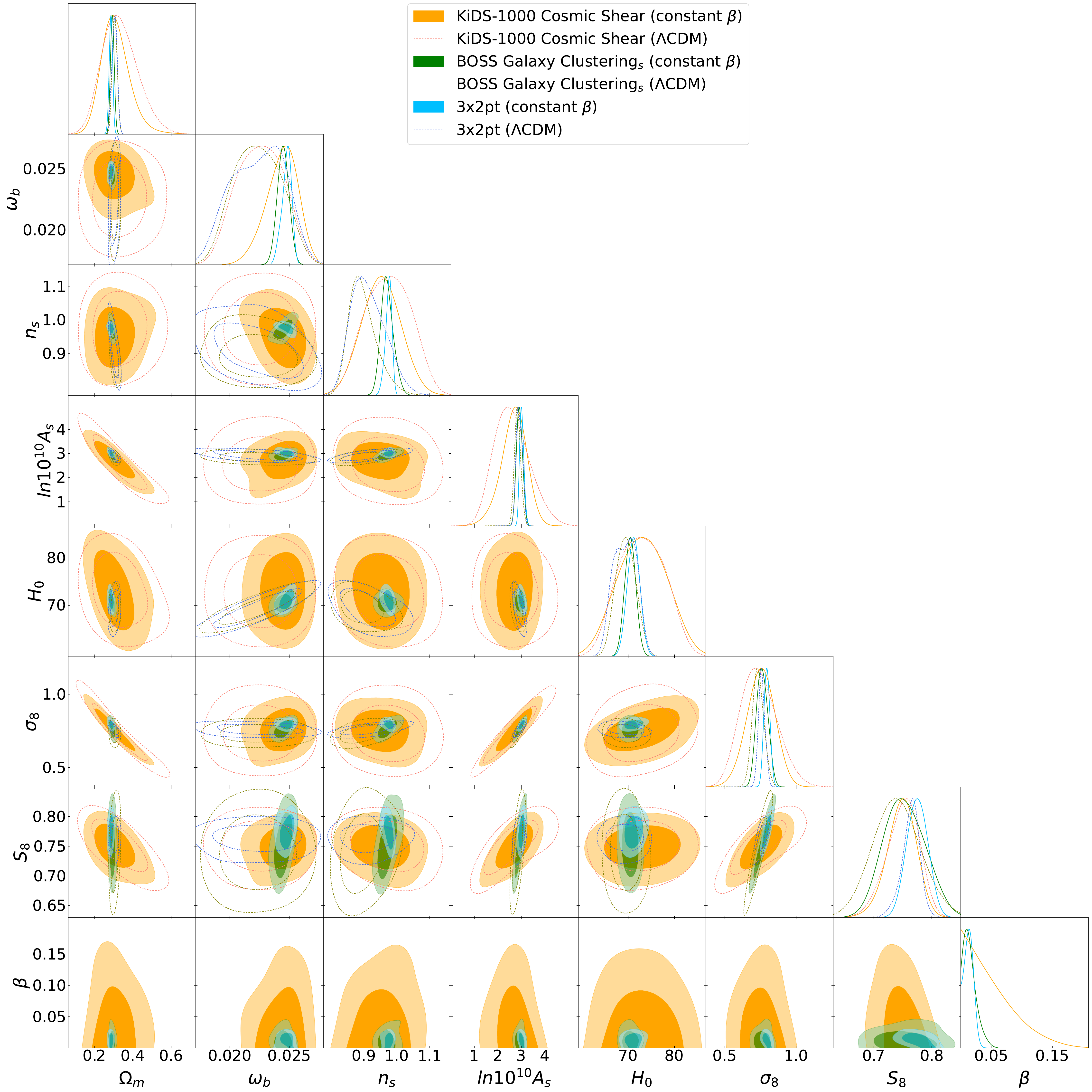}
    \caption{Triangular plot of 68\% and 95\% C.L. posterior distributions of the various cosmological parameters, using datasets: KiDS-1000 cosmic shear (orange), BOSS spectroscopic galaxy clustering (green) and their 3x2pt (blue), for the $\Lambda$CDM (unfilled contours) and the constant coupling model. }
    \label{fig:full_WL_contour_const}
\end{figure*}

Finally, we consider constraints from galaxy surveys, using cosmic shear, galaxy clustering and 3x2pt: we use these datasets here for the first time in the context of CDE cosmologies, both for a constant coupling and for a tomographic coupling. The case of a constant coupling is shown in Fig.~\ref{fig:full_WL_contour_const}: $\beta$ is mainly constrained by galaxy clustering and 3x2pt probes, rather than by cosmic shear from KiDS alone (at least for the conservative cut we use at non-linear scales, cf. Sec. \ref{sec:WLdata}). 

\begin{figure*}[t!]
    \centering
    \includegraphics[width=\linewidth]{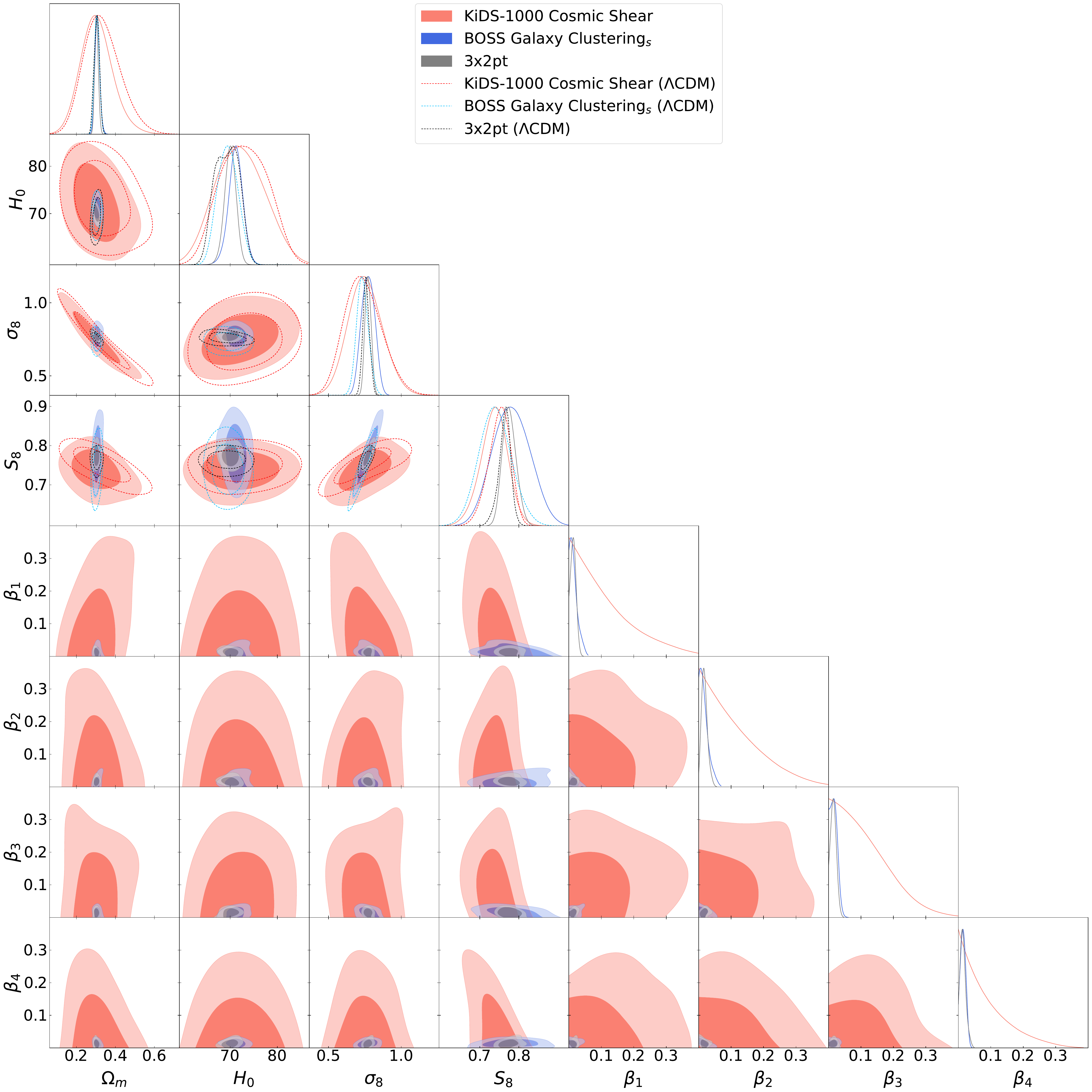}
    \caption{Triangular plot of 68\% and 95\% C.L. posterior distributions of the various cosmological parameters, for both $\Lambda$CDM and the 4-bin tomographic model using datasets: KiDS-1000 cosmic shear (red), BOSS spectroscopic galaxy clustering (blue) and their 3x2pt (grey). For reference, the bin edges are defined as $z=\{0,0.5,1,2\}$.}
    \label{fig:full_WL_contour}
\end{figure*}

\begin{figure}[t!]
            \centering    
            \includegraphics[width=\linewidth]{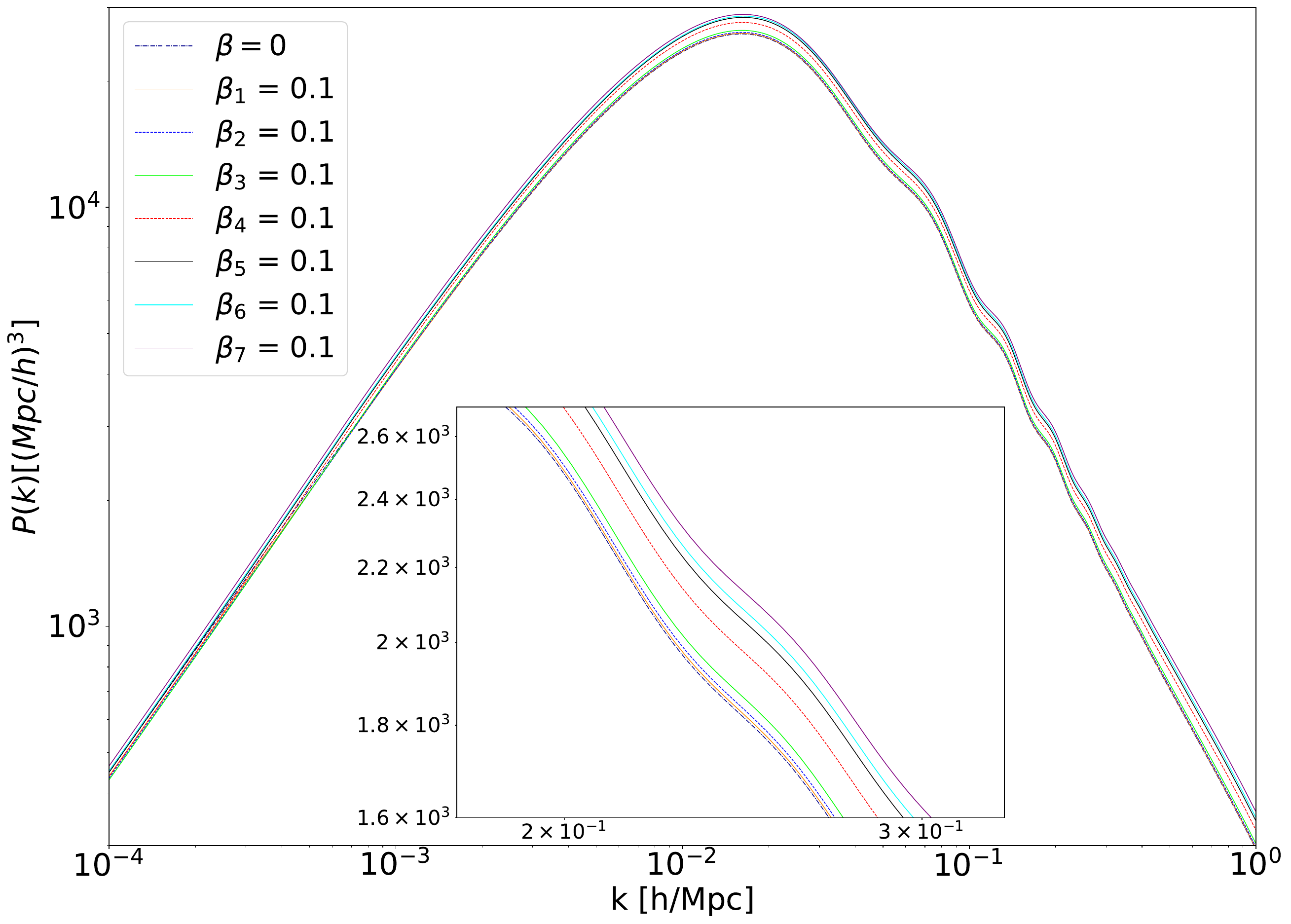}
         \caption
        {Plot of the matter power spectrum $P(k)$ when activating coupling at each redshift bin (increasing subsequent values of coupling from 0 to 0.1 as done in Fig.~\ref{fig:varying_beta}), in the tomographic 7-bin case, keeping initial conditions fixed. The dashed dark blue line with $\beta=0$ corresponds to $\Lambda$CDM. The inset plot is a zoom-in at small $k$ scales. The binning is defined by edges $z=\{0,1,2,5,100,500,1000\}$.} 
        \label{fig:7-bin-Pk}
\end{figure}

Next, we present results using the same probes, but for a 4-bin tomographic coupling, in Fig.~\ref{fig:full_WL_contour}. We see that once again, similar to the constant coupling case, coupling strength is relatively unconstrained when using cosmic shear from KiDS alone (with a conservative cut),  even more so than in the constant coupling case. Also, coupling at the highest redshift bin, $\beta_4$, is more constrained than the lower redshift bins. The lack of constraining power at $z < 2$ could be due to the fact that, as shown in Fig.~\ref{fig:7-bin-Pk}, the increase of $P(k)$ caused by the coupling at such low redshifts is too small for cosmic shear to be sensitive towards. Moreover, cosmic shear is able to constrain $S_8$ but exhibits a strong degeneracy in the $\sigma_8-\Omega_m$ plane, which works in the same direction as $\beta$: larger values of $\beta$ increase $\sigma_8$ while decreasing $\Omega_m$. This also explains why larger values of the coupling are allowed by this dataset. On the other hand, galaxy clustering and subsequently 3x2pt are able to give tight constraints on the coupling at all redshifts, as they are able to break this degeneracy. 


\begin{figure}[t!]
    \centering
    \includegraphics[width=\linewidth]{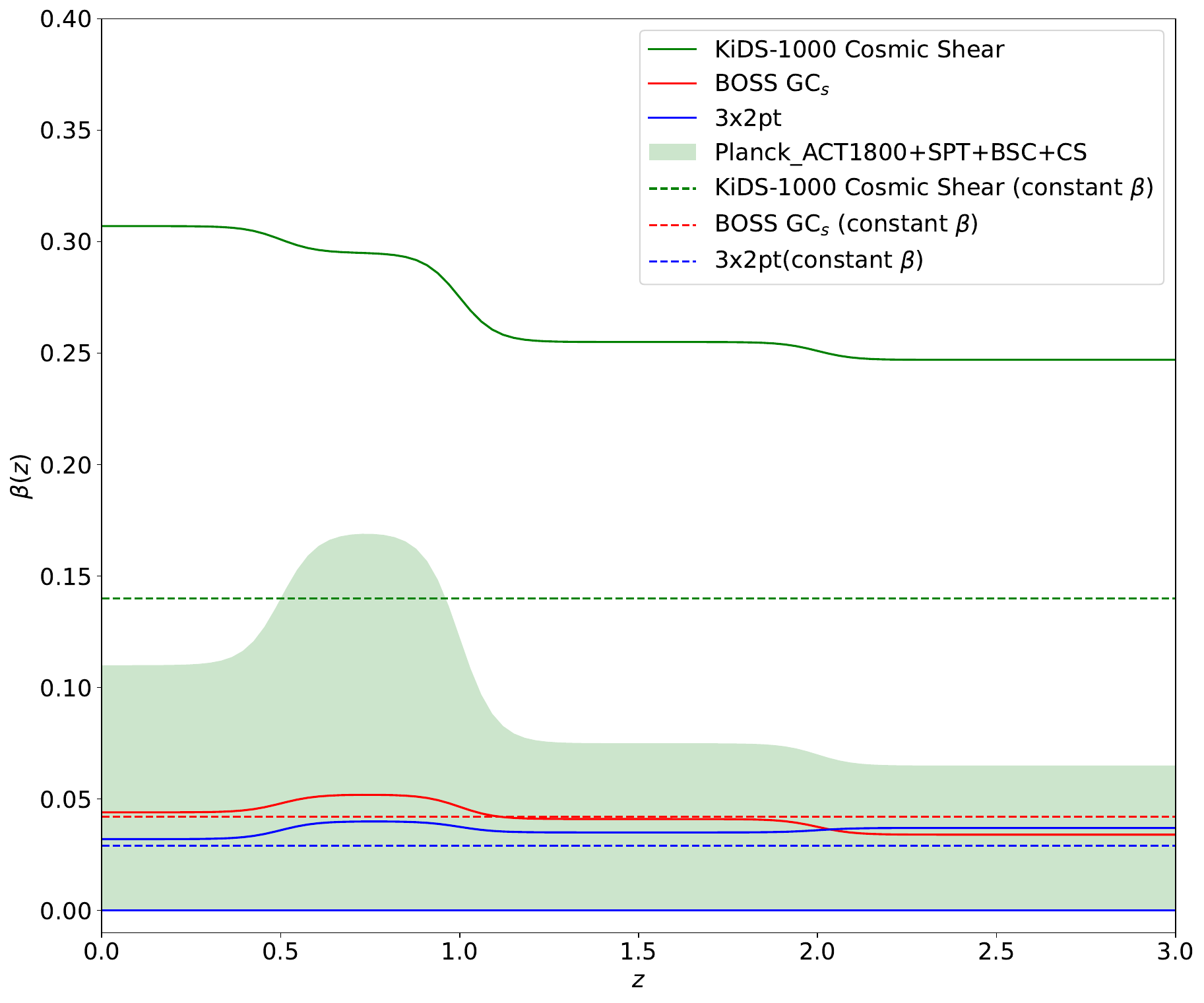}
    \caption{\small 95\% C.L. on $\beta(z)$ for a 4-bin coupling model, for cosmic shear (green), galaxy clustering (red) and 3x2pt (blue) datasets, tested on CDE with a constant coupling (dotted lines) and tomographic CDE (solid lines). The Planck+ACT1800+SPT+BSC+CS dataset, tested with a tomographic CDE model, is shaded in green. The binning is defined by edges $z=\{0,0.5,1,2\}$. }
    \label{fig:beta_lim_WL}
\end{figure}

In Fig.~\ref{fig:beta_lim_WL} we plot the reconstruction of $\beta(z)$ at 95\% C.L. for the various datasets we have employed. We see that when considering only the weak lensing probe, $\beta(z)$ is once again less constrained at low redshifts. However, when Planck+ACT1800+SPT+BSC data is included, constraints are tightened significantly (going from CS to the Planck+ACT1800+SPT+BSC+CS dataset). Moreover, comparing between CDE models, we see that although constraints become much weaker going from a constant to a tomographic coupling model in the case of cosmic shear, constraints on $\beta$ remain very similar for GC and 3x2pt. Remarkably, GC and 3x2pt exhibit comparable, or even stronger, constraining power compared to CMB probes. Hence weak lensing and galaxy clustering are powerful tools to constrain coupled dark energy models, as they are sensitive to large-scale structure formation during late times and  give considerable constraints on coupling at higher redshifts. 

\begin{figure*}[ht]
    \centering
    \includegraphics[width=\linewidth]{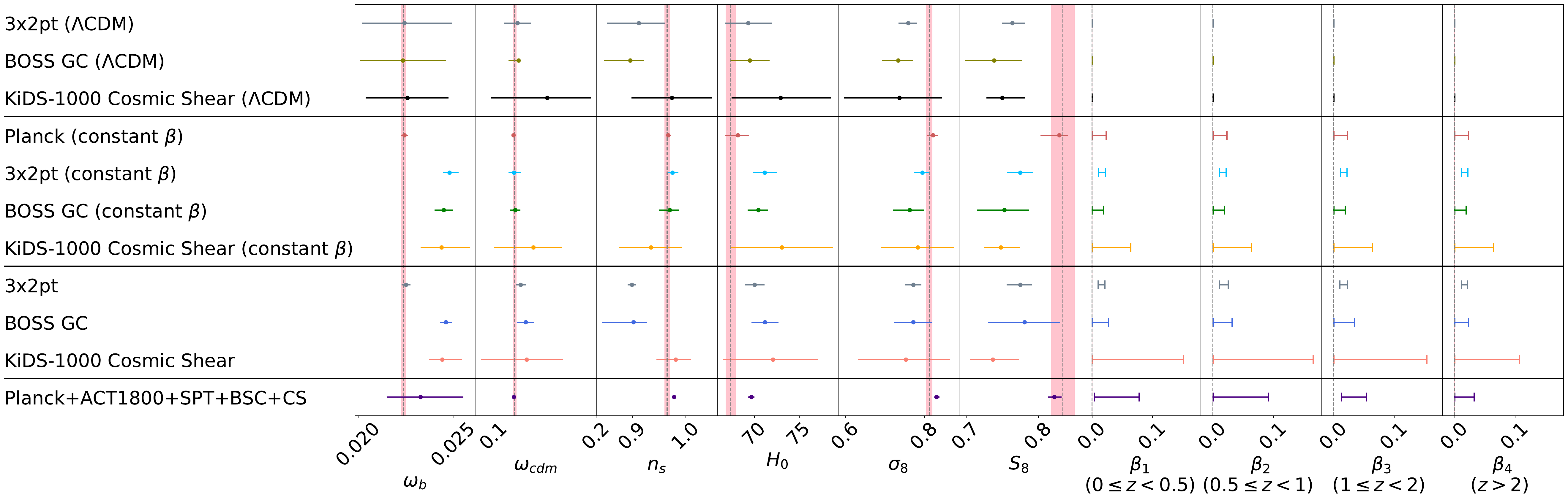}
    \caption{Mean and 68\% C.L. values for the various cosmological parameters and coupling coefficients obtained with the three weak lensing and galaxy clustering datasets only, for the $\Lambda$CDM, the $\beta$=const. and the 4-bin tomographic models. For ease of comparison, we put in the same figure results obtained from our earlier analysis of a Planck dataset in a constant coupling case, and results }from the analysis of Planck+ACT1800+SPT+BSC+CS in a 4-bin tomographic model.
    \label{fig:horiz_bestfit_WL}
\end{figure*}

In Fig.~\ref{fig:horiz_bestfit_WL} we plot the mean and $68\%$ C.L. of the full set of cosmological parameters for each dataset, comparing them to the results obtained assuming $\Lambda$CDM. We see that in the case of a combined CMB and cosmic shear dataset, the bulk of the statistical power comes from Planck data: the mean value of each parameter is now more in agreement with $\textit{Planck}$ fiducial cosmology. Moreover, with a CDE model, the tension between the measurement of the $S_8$ parameter from CMB and weak lensing is eased: for $\Lambda$CDM, $S_8=0.764^{+0.017}_{-0.014}$ with the 3x2pt probe, and $S_8=0.834\pm0.016$ with the Planck dataset, which is a ${\sim} 3.1\sigma$ tension. In a constant coupling model, $S_8=0.775\pm0.018$ using 3x2pt and $S_8 = 0.829^{+0.023}_{-0.026}$ with Planck, thus reducing this tension to ${\sim} 1.8\sigma$. The decrease of the tension is due to the increase of $S_8$ obtained from the 3x2pt dataset and also to the increase of the uncertainties. Both effects are induced by the coupling. On the other hand, the values for the other parameters remain within $1\sigma$ agreement: for example, in a constant coupling case, $\omega_\mathrm{cdm}=0.119\pm 0.002$ and $\sigma_8=0.822^{+0.013}_{-0.015}$ for the Planck dataset, while $\omega_\mathrm{cdm}=0.120^{+0.007}_{-0.006}$ and $\sigma_8=0.795^{+0.020}_{-0.015}$ for the 3x2pt dataset. However, we note that these calculations are only estimates as they assume Gaussian posterior distributions for the cosmological parameters.

These calculations have been made based on the values  quoted in Tables~\ref{table:cmb_bestfit} and \ref{table:WL_bestfit} of Appendix \ref{sec:AppendixA}. We also include the best-fit, mean and 68\% C.L. uncertainties for each parameter.


\section{Model Comparison}\label{sec:model-comparison}
We conduct a rough model comparison by calculating the $\Delta \chi^2_\mathrm{min}= \chi^2_\mathrm{CDE,min}-\chi^2_{\Lambda\mathrm{CDM,min}}$ of each of the datasets used in each model with respect to $\Lambda$CDM, and report their values in Tables~\ref{table:cmb_bestfit}-\ref{table:planck_WL_bestfit}. We see that across all models and datasets, the CDE model (be it constant or tomographic) is able to reduce the $\chi^2_\mathrm{min}$ value with respect to $\Lambda$CDM. This difference is generally not large enough to signify a statistically significant preference for CDE over $\Lambda$CDM; moreover, the large number of parameters varied especially in the tomographic CDE scenario will penalise the model.

Thus we also calculate the Bayes' ratio, a more robust method of quantifying the goodness of fit of the data to the model as it takes into account the number of parameters varied in the model. Its logarithm is given by $\ln{B}=\ln{E(\mathrm{CDE}|D)}-\ln{E(\Lambda\mathrm{CDM}|D)}$, where $E(M_i|D)$ is the Bayesian evidence of the model $M_i$ under the dataset $D$, see e.g. \cite{Trotta:2017wnx}. We use the code $\texttt{MCEvidence}$ \cite{Heavens:2017afc} to numerically calculate the evidence estimated using the kth Nearest Neighbour method, for one chosen dataset in each CDE model (3-bin, 7-bin, 4-bin and constant coupling when using cosmic shear). We report their values in Tables~\ref{table:cmb_bestfit}-\ref{table:planck_WL_bestfit}. We see that $\ln{B}<-3$ for all models which, based on the Jeffreys' scale, shows substantial evidence for $\Lambda$CDM over CDE models \cite{Kass:1995loi}. Thus even when constant or tomographic CDE models give a smaller $\chi^2_\mathrm{min}$, the data does not show a preference towards CDE.

\section{Conclusions}\label{sec:conclusions}
We have introduced a new parameterisation for a tomographic coupled dark energy model, where we allow the coupling strength between the dark energy scalar field and dark matter to vary within redshift bins. To constrain our model, we first used several combinations of CMB data to understand how different scale cuts could affect the coupling constraints. Subsequently, we added low redshift datasets successively, considering tomographic CDE models with finer binning at low redshifts, to test the constraining power of each dataset. Hence we defined three different binning models: 1) a 3-bin model with bin edges $z=\{0,100,1000\}$, where we primarily investigated the capability of state-of-the-art CMB data from {\it Planck}, ACT and SPT to constrain the coupling before and after recombination; 2) a 7-bin model more finely binned at low redshifts, with edges $z=\{0,1,2,5,100,500,1000\}$ probed by a combination of CMB and low-$z$ background and large-scale structure data; and 3) a 4-bin model only varying at low redshifts, with bin edges $z=\{0,0.5,1,2\}$, where we exploit for the first time in the context of CDE the weak lensing (cosmic shear) data from KiDS, the spectroscopic full-shape galaxy clustering data from BOSS, and the 3x2pt likelihood, which includes their cross-correlation. 

In the case of a 3-bin model probed with early-time CMB data, we see that constraints on $\beta$ are larger for a tomographic case as compared to in a constant coupling model. The Planck dataset gives the tightest constraints, with the amplitude of coupling at pre-recombination eras $\beta_3 < 0.026$ at 68\% C.L., increasing slightly to $\beta_1 < 0.031$ at later times. The inclusion of ACT and SPT data loosens these constraints, with $\beta_3$ going up to $\beta_3 < 0.066$ for a Planck+ACT1800+SPT dataset. When we perform a cut on Planck data (i.e. in the Planck650+ACT+SPT dataset), we obtain, as expected, the loosest constraints on coupling, with $\beta_3 < 0.068$.

Next, when we introduce background probes and employ a finer 7-binning model, we find that constraints on coupling once again loosen, especially at low redshifts. Taking the example of a Planck+ACT1800+SPT+BSC dataset, $\beta = 0.015^{+0.005}_{-0.014}$ in a constant case, but when a constant assumption is relaxed, $\beta_1=0.146^{+0.075}_{-0.081}$ when $z<1$, which is almost an order of magnitude increase in coupling. Hence through this study, we have seen that coupling strength is highly sensitive to the different epochs in the history of the Universe, a feature that we have been able to probe by using different datasets. Additionally, the tightest constraints on coupling are found in the range $z\in[100,500]$, where $\beta_5<0.028$ at 68\% C.L. with Planck+ACT1800+SPT+BSC. When we add RSD we get stronger constraints in all bins. On the other hand, a SH0ES prior increases the 68\% upper bound on $\beta_5$ to 0.061. When we compare the 2 baseline CMB datasets we have employed (namely a Planck+ACT1800+SPT cut versus a Planck650+ACT+SPT cut) we see that they can affect coupling constraints at high redshifts, with the Planck650 cut allowing for larger values of $\beta_7$. These constraints are captured in Fig.~\ref{fig:horiz_bestfit_planckcut}, where we can see how constraints on the $\beta$ coefficients and the various cosmological parameters, including $H_0$ and $S_8$, vary with the various datasets we have tested. Moreover, a tomographic coupling, which leads to an increase in clustering and a faster decay of cold dark matter as compared to $\Lambda$CDM, leads to generally larger values of $H_0$ and $\sigma_8$ for all datasets, when compared to the $\textit{Planck}$ fiducial $\Lambda$CDM cosmology. 

We have also, for the first time, utilised the cross-correlation between weak lensing and full-shape galaxy clustering probes to test CDE cosmologies for both a constant and tomographic coupling case. We see that spectroscopic galaxy clustering and 3x2pt probes can provide very tight constraints on coupling strength even for a tomographic case. Taking the example redshift range $0.5 < z < 1$, $\beta_2 = 0.018^{+0.007}_{-0.011}$, bringing the 3x2pt probe on the same level of constraining power as CMB and background datasets when employed in a constant coupling case. Thus, it becomes a powerful tool to constrain such CDE models, with independent systematics from high-redshift probes like the CMB.

In terms of $H_0$ and $S_8$, we see that within the CDE framework, the tensions are smaller: the $H_0$ tension decreases from ${\sim} 4.8\sigma$ (assuming $\Lambda$CDM) to ${\sim} 2.9\sigma$ (assuming constant coupling CDE) if we take the results obtained with the Planck dataset and compare them to the $H_0$ value as quoted by SH0ES. In the case of the tomographic CDE model, the $H_0$ tension decreases even more, to ${\sim} 2.7\sigma$. For $S_8$, the tension is alleviated to an even larger extent: from ${\sim} 3.1 \sigma$ in $\Lambda$CDM (when comparing between Planck and 3x2pt datasets) to ${\sim}1.8\sigma$ in a constant coupling model. We note, however, that this reduction in the tension is mainly due to the increase in the uncertainties of $H_0$ and $S_8$ that we have derived with a CDE model, rather than to a shift in their mean value.

We do not find evidence for non-zero coupling, either for a constant or for a tomographic case. A non-zero coupling is however still in agreement with current data. It is still worth reiterating that all the datasets we have employed still favour $\Lambda$CDM, when the Bayes' ratio is used for model comparison. With respect to  \cite{Gomez-Valent:2020mqn},
we have updated constraints with new CMB datasets from ACT and SPT; we have used for the first time galaxy-galaxy lensing to probe a tomographic CDE model, demonstrating how important this probe is to test the coupling at small redshifts.

Finally, we note that, in absence of simulations for varying coupling, we did the conservative choice of cutting most of the non-linear scales; to fully exploit the constraining power of low redshift datasets on a CDE model, it is essential to have more accurate calculations for the nonlinear matter power spectrum, especially in the case of a tomographic CDE model. With upcoming CMB surveys such as the CMB-S4 \cite{CMB-S4:2020lpa} and LiteBIRD \cite{LiteBIRD:2020khw}, as well as Stage IV weak lensing and galaxy clustering surveys like $\textit{Euclid}$ \cite{EUCLID:2011zbd}, the Legacy Survey of Space and Time (LSST) \cite{LSST:2008ijt} and the Dark Energy Spectroscopic Instrument (DESI) \cite{DESI:2016fyo}, more accurate and precise data will become available to test the strength of fifth force interactions in the dark sector and their role in the background expansion and the formation of large-scale structures in the universe, potentially leading to a non-zero coupling detection.


\vspace{1.25cm}
{\bf Acknowledgements}
\newline
\newline
The authors would like to thank Luca Amendola, Christof Wetterich and Natalie Hogg for their insightful comments and discussions. AGV is funded by the Instituto Nazionale di Fisica Nucleare (INFN) through the project of the InDark INFN Special Initiative: ``Dark Energy and Modified Gravity Models in the light of Low-Redshift Observations'' (n. 22425/2020). He is grateful to CEA Paris-Saclay for the kind hospitality during his visit in June 2022.
\vspace{1.25cm}
\appendix
\section{Best Fit Cosmological Parameters}\label{sec:AppendixA}
Here we present the tables with the best-fit, mean and 68\% C.L. values for the various models we have tested.

\begin{table*}
\centering
\scriptsize
\begin{tabular}{| *{13}{c|} }
 \hline
\multicolumn{13}{|c|}{3-bin Tomographic $\beta$}\\
                    \hline
            & \multicolumn{2}{c|}{\textbf{Planck} }
             & \multicolumn{2}{c|}{\textbf{Planck} }
            & \multicolumn{2}{c|}{\textbf{Planck}}
            & \multicolumn{2}{c|}{\textbf{Planck}}
                    & \multicolumn{2}{c|}{\textbf{Planck}}
                    & \multicolumn{2}{c|}{\textbf{Planck650}}\\
                    &\multicolumn{2}{c|}{(constant $\beta$)}
                    &\multicolumn{2}{c|}{}
                    &\multicolumn{2}{c|}{\textbf{+PlanckLens}}
                    & \multicolumn{2}{c|}{\textbf{+ACT1800}}
                    & \multicolumn{2}{c|}{\textbf{+ACT1800+SPT}}
                    & \multicolumn{2}{c|}{\textbf{+ACT+SPT}}
                    \\
                    \hline
Parameter & Best-Fit & Mean$\pm\sigma$& Best-Fit & Mean$\pm\sigma$& Best-Fit & Mean$\pm\sigma$& Best-Fit & Mean$\pm\sigma$& Best-Fit & Mean$\pm\sigma$& Best-Fit & Mean$\pm\sigma$\\ \hline 
$100~\omega{}_{\rm b}$& $2.248$ & $2.241 _{-0.019}^{+0.018}$& $2.235$ & $2.252 _{-0.020}^{+0.020}$& $2.256$ & $2.244 _{-0.023}^{+0.022}$& $2.247$ & $2.246 _{-0.022}^{+0.020}$& $2.258$ & $2.245 _{-0.014}^{+0.012}$& $2.229$ & $2.250 \pm0.014$\\
$\omega{}_{\rm cdm }$& $0.118$ & $0.119 \pm 0.002$& $0.121$ & $0.120 _{-0.004}^{+0.003}$& $0.119$ & $0.120 _{-0.004}^{+0.003}$& $0.119$ & $0.119 _{-0.004}^{+0.003}$& $0.116$ & $0.119 _{-0.001}^{+0.002}$& $0.120$ & $0.118 _{-0.002}^{+0.002}$\\
$\ln{10^{10}A_{s }}$& $3.042$ & $3.048 _{-0.015}^{+0.014}$& $3.046$ & $3.049 _{-0.021}^{+0.018}$& $3.051$ & $3.051 _{-0.026}^{+0.026}$& $3.069$ & $3.057 _{-0.022}^{+0.021}$& $3.052$ & $3.056 _{-0.014}^{+0.012}$& $3.043$ & $3.047 _{-0.014}^{+0.012}$\\
$n_{\rm s }$& $0.970$ & $0.967\pm 0.005$& $0.966$ & $0.971\pm 0.007$& $0.972$ & $0.971 _{-0.004}^{+0.005}$& $0.974$ & $0.971\pm 0.005$& $0.973$ & $0.973 \pm 0.004$& $0.976$ & $0.977\pm 0.005$\\
$\tau{}_{\rm reio }$& $0.052$ & $0.056 _{-0.009}^{+0.008}$& $0.053$ & $0.057 _{-0.012}^{+0.009}$& $0.054$ & $0.057 _{-0.013}^{+0.012}$& $0.058$ & $0.057 _{-0.009}^{+0.008}$& $0.053$ & $0.056 _{-0.008}^{+0.007}$& $0.051$ & $0.055 \pm 0.006$\\
$\sigma_8$& $0.829$ & $0.822 _{-0.015}^{+0.013}$& $0.820$ & $0.825 _{-0.017}^{+0.014}$& $0.813$ & $0.817 _{-0.015}^{+0.013}$& $0.86$ & $0.837 _{-0.028}^{+0.023}$& $0.841$ & $0.832 _{-0.014}^{+0.011}$& $0.819$ & $0.814 _{-0.009}^{+0.008}$\\
$S_8$& $0.819$ & $0.829 _{-0.026}^{+0.023}$& $0.839$ & $0.831 _{-0.043}^{+0.033}$& $0.821$ & $0.828 _{-0.026}^{+0.018}$& $0.843$ & $0.832 _{-0.033}^{+0.022}$& $0.811$ & $0.828 _{-0.016}^{+0.014}$& $0.842$ & $0.8193\pm 0.018$\\
$H_0$& $69.17$ & $68.15 _{-1.43}^{+1.21}$& $67.62$ & $68.51 _{-1.17}^{+1.42}$& $68.02$ & $67.93 _{-0.77}^{+0.97}$& $69.89$ & $69.17 _{-1.65}^{+1.75}$& $70.48$ & $68.98 _{-1.03}^{+0.79}$& $66.99$ & $68.07 _{-0.71}^{+0.63}$\\
$\beta_1$& $0.03$ & $<0.023$& $0.018$ & $<0.031$& $0.0$ & $<0.014$& $0.027$ & $<0.042$& $0.057$ & $<0.033$& $0.031$ & $<0.059$\\
$\beta_2$& $-$ & $-$& $0.014$ & $<0.029$& $0.004$ & $<0.027$& $0.03$ & $<0.037$& $0.036$ & $<0.029$& $0.042$ & $<0.042$\\
$\beta_3$& $-$ & $-$& $0.02$ & $<0.026$& $0.024$ & $<0.024$& $0.112$ & $<0.036$& $0.066$ & $<0.066$& $0.071$ & $<0.068$\\
\hline
$\Delta \chi^2$& $-2.59$ && $-1.99$ && $-1.72$&&$-1.12$&&$-5.96$&&$-6.11$& \\
$\ln{B}$&&& $-3.73$ && &&&&&&&\\
\hline 
 \end{tabular}   
 \caption{Table of best-fit, mean and 68\% C.L.  values of the various cosmological parameters of the 3-bin tomographic model, for the CMB datasets Planck, Planck+ACT1800, Planck+ACT1800+SPT and Planck650+ACT+SPT. We also include the case of a constant coupling model using Planck data. For reference, the binning is defined by edges $z=\{0,100,1000\}$. In the last two rows, we include the $\Delta \chi^2 = \chi^2_\mathrm{min,CDE}-\chi^2_\mathrm{min,\Lambda CDM}$ difference for each model (constant and tomographic CDE) and dataset, compared with $\Lambda$CDM using the same dataset, as well as the logarithm of the Bayes' ratio $\ln{B}=\ln{E_\mathrm{CDE}}-\ln{E_{\Lambda \mathrm{CDM}}}$ for one chosen dataset. }
 \label{table:cmb_bestfit}
\end{table*}

\begin{table*}
\centering
\footnotesize
\begin{tabular}{| *{11}{c|} }
 \hline
\multicolumn{11}{|c|}{7-bin Tomographic $\beta$}\\
                    \hline
            & \multicolumn{2}{c|}{\textbf{Planck+ACT1800} }
            & \multicolumn{2}{c|}{\textbf{Planck+ACT1800} }
            & \multicolumn{2}{c|}{\textbf{Planck+ACT1800}}
                    & \multicolumn{2}{c|}{\textbf{Planck+ACT1800}}
                    & \multicolumn{2}{c|}{\textbf{Planck+ACT1800}}\\
                    &\multicolumn{2}{c|}{\textbf{+SPT+BSC }}
                    &\multicolumn{2}{c|}{\textbf{+SPT+BSC}}
                    & \multicolumn{2}{c|}{\textbf{+SPT+BSC}}
                    & \multicolumn{2}{c|}{\textbf{+SPT+BSC}}
                    & \multicolumn{2}{c|}{\textbf{+SPT+BSC}}
                    \\
                    &\multicolumn{2}{c|}{\textbf{(constant $\beta$)}}
                    &\multicolumn{2}{c|}{\textbf{}}
                    & \multicolumn{2}{c|}{\textbf{RSD}}
                    & \multicolumn{2}{c|}{\textbf{SH0ES}}
                    & \multicolumn{2}{c|}{\textbf{RSD+SH0ES}}
                    \\
                    \hline
Parameter & Best-Fit & Mean$\pm\sigma$& Best-Fit & Mean$\pm\sigma$& Best-Fit & Mean$\pm\sigma$& Best-Fit & Mean$\pm\sigma$& Best-Fit & Mean$\pm\sigma$ \\ \hline 
$100~\omega_\mathrm{b}$& $2.235$ & $2.244 _{-0.016}^{+0.012}$& $2.254$ & $2.239\pm 0.014$& $2.239$ & $2.246 _{-0.013}^{+0.014}$& $2.235$ & $2.241 _{-0.010}^{+0.012}$& $2.252$ & $2.247 _{-0.013}^{+0.011}$\\
$\omega_\mathrm{cdm }$& $0.117$ & $0.1178 _{-0.0008}^{+0.0012}$& $0.1172$ & $0.116 _{-0.0015}^{+0.0022}$& $0.1167$ & $0.1169 _{-0.0009}^{+0.0014}$& $0.1154$ & $0.1167\pm 0.0008$& $0.1162$ & $0.1159\pm 0.0011$\\
$\ln{10^{10}A_\mathrm{s}}$& $3.047$ & $3.055 _{-0.017}^{+0.015}$& $3.047$ & $3.052 _{-0.015}^{+0.014}$& $3.030$ & $3.050 _{-0.015}^{+0.013}$& $3.046$ & $3.052 _{-0.015}^{+0.013}$& $3.050$ & $3.044 _{-0.020}^{+0.015}$\\
$n_\mathrm{s}$& $0.976$ & $0.971 _{-0.004}^{+0.005}$& $0.972$ & $0.970 _{-0.004}^{+0.005}$& $0.974$ & $0.972\pm 0.005$& $0.969$ & $0.969 \pm 0.003$& $0.974$ & $0.972 _{-0.003}^{+0.004}$\\
$\tau_\mathrm{reio }$& $0.055$ & $0.058 _{-0.009}^{+0.006}$& $0.054$ & $0.054 _{-0.008}^{+0.007}$& $0.05$ & $0.057 _{-0.006}^{+0.007}$& $0.054$ & $0.055 \pm 0.007$& $0.057$ & $0.053 _{-0.009}^{+0.007}$\\
$\sigma_8$& $0.821$ & $0.818 _{-0.011}^{+0.010}$& $0.832$ & $0.845 _{-0.022}^{+0.013}$& $0.807$ & $0.822 _{-0.011}^{+0.003}$& $0.83$ & $0.834 _{-0.010}^{+0.010}$& $0.817$ & $0.828 _{-0.014}^{+0.012}$\\
$S_8$& $0.815$ & $0.817 _{-0.010}^{+0.014}$& $0.832$ & $0.834 _{-0.017}^{+0.015}$& $0.801$ & $0.814 _{-0.011}^{+0.010}$& $0.803$ & $0.818 _{-0.011}^{+0.010}$& $0.802$ & $0.809 _{-0.013}^{+0.011}$\\
$H_0$& $68.73$ & $68.49 _{-0.56}^{+0.46}$& $68.27$ & $68.86 \pm 0.59$& $68.51$ & $68.85 _{-0.60}^{+0.38}$& $70.06$ & $69.5 _{-0.43}^{+0.45}$& $69.18$ & $69.51 _{-0.55}^{+0.46}$\\
$\beta_1$& $0.023$ & $0.015 _{-0.014}^{+0.005}$& $0.185$ & $0.146 _{-0.081}^{+0.075}$& $0.03$ & $0.027 _{-0.026}^{+0.005}$& $0.014$ & $<0.017$& $0.01$ & $<0.038$\\
$\beta_2$& $-$ & $-$& $0.036$ & $<0.124$& $0.057$ & $<0.015$& $0.033$ & $0.028 _{-0.023}^{+0.010}$& $0.088$ & $<0.038$\\
$\beta_3$& $-$ & $-$& $0.078$ & $<0.094$& $0.012$ & $0.064 _{-0.023}^{+0.018}$& $0.019$ & $0.041 _{-0.016}^{+0.016}$& $0.051$ & $<0.048$\\
$\beta_4$&$-$ & $-$& $0.024$ & $<0.031$& $0.031$ & $<0.016$& $0.067$ & $0.058 _{-0.015}^{+0.018}$& $0.03$ & $<0.041$\\
$\beta_5$& $-$ & $-$& $0.005$ & $<0.028$& $0.054$ & $0.031 _{-0.011}^{+0.010}$& $0.055$ & $0.050 _{-0.013}^{+0.011}$& $0.008$ & $<0.048$\\
$\beta_6$& $-$ & $-$& $0.029$ & $<0.054$& $0.011$ & $0.030 _{-0.019}^{+0.011}$& $0.022$ & $<0.027$& $0.048$ & $<0.049$\\
$\beta_7$& $-$ & $-$& $0.059$ & $0.063 _{-0.054}^{+0.023}$& $0.001$ & $<0.018$& $0.006$ & $0.022 _{-0.014}^{+0.010}$& $0.03$ & $<0.042$\\
\hline
$\Delta \chi^2$& $-0.80$ && $-2.66$ && $-2.56$&&$-9.88$&&$-1.92$& \\
$\ln{B}$&&& $-10.61$ && &&&&& \\
\hline 
 \end{tabular}   
 \caption{Table of best-fit, mean and 68\% C.L.  values of the various cosmological parameters of the 7-bin tomographic model, using the baseline CMB dataset with a cut in ACT, i.e. Planck+ACT1800+SPT, in combination with BSC, BSC+RSD, BSC+SH0ES or BSC+RSD+SH0ES. The binning is defined in this case by edges $z=\{0,1,2,5,100,500,1000\}$. In the last two rows, we include the $\Delta \chi^2 = \chi^2_\mathrm{min,CDE}-\chi^2_\mathrm{min,\Lambda CDM}$ difference for each model (constant and tomographic CDE) and dataset, compared with $\Lambda$CDM using the same dataset, as well as the logarithm of the Bayes' ratio $\ln{B}=\ln{E_\mathrm{CDE}}-\ln{E_{\Lambda \mathrm{CDM}}}$ for one chosen dataset.}
 \label{table:actpolcut_bestfit}
\end{table*}

\begin{table*}
\centering
\begin{tabular}{| *{9}{c|} }
 \hline
  \multicolumn{9}{|c|}{7-bin Tomographic $\beta$}\\
                    \hline
            & \multicolumn{2}{c|}{\textbf{Planck650+ACT} }
            & \multicolumn{2}{c|}{\textbf{Planck650+ACT} }
            & \multicolumn{2}{c|}{\textbf{Planck650+ACT}}
                    & \multicolumn{2}{c|}{\textbf{Planck650+ACT}}\\
                    &\multicolumn{2}{c|}{\textbf{+SPT+BSC (constant $\beta$)}}
                    &\multicolumn{2}{c|}{\textbf{+SPT+BSC}}
                    & \multicolumn{2}{c|}{\textbf{+SPT+BSC+RSD}}
                    & \multicolumn{2}{c|}{\textbf{+SPT+BSC+SH0ES}}
                    \\
                    \hline
Parameter & Best-Fit & Mean$\pm\sigma$& Best-Fit & Mean$\pm\sigma$& Best-Fit & Mean$\pm\sigma$& Best-Fit & Mean$\pm\sigma$\\ \hline 
$100~\omega_\mathrm{b}$& $2.263$ & $2.252 _{-0.012}^{+0.011}$& $2.251$ & $2.247 \pm 0.012$& $2.252$ & $2.255 _{-0.011}^{+0.012}$& $2.245$ & $2.239 _{-0.015}^{+0.017}$\\
$\omega{}_\mathrm{cdm }$& $0.1175$ & $0.1178\pm 0.0009$& $0.1172$ & $0.1157 _{-0.0013}^{+0.0028}$& $0.1171$ & $0.1165 _{-0.0008}^{+0.0010}$& $0.1173$ & $0.1153 _{-0.0019}^{+0.0029}$\\
$\ln{10^{10}A_\mathrm{s }}$& $3.05$ & $3.052\pm 0.005$& $3.049$ & $3.048 _{-0.013}^{+0.011}$& $3.036$ & $3.035 _{-0.012}^{+0.010}$& $3.064$ & $3.059 _{-0.018}^{+0.013}$\\
$n_\mathrm{s }$& $0.977$ & $0.976\pm 0.005$& $0.975$ & $0.977\pm 0.008$& $0.978$ & $0.978\pm 0.004$& $0.973$ & $0.971\pm 0.004$\\
$\tau{}_\mathrm{reio }$& $0.058$ & $0.057 _{-0.003}^{+0.004}$& $0.054$ & $0.054 \pm 0.006$& $0.051$ & $0.050 _{-0.006}^{+0.005}$& $0.057$ & $0.057 _{-0.008}^{+0.006}$\\
$\sigma_8$& $0.808$ & $0.816 _{-0.009}^{+0.003}$& $0.834$ & $0.844 _{-0.022}^{+0.015}$& $0.808$ & $0.813 _{-0.009}^{+0.007}$& $0.868$ & $0.870 _{-0.032}^{+0.020}$\\
$S_8$& $0.805$ & $0.815 \pm 0.010$& $0.828$ & $0.829 _{-0.017}^{+0.013}$& $0.802$ & $0.803\pm 0.01$& $0.848$ & $0.842 _{-0.027}^{+0.015}$\\
$H_0$& $68.57$ & $68.47 _{-0.51}^{+0.45}$& $68.76$ & $69.10 _{-0.51}^{+0.46}$& $68.67$ & $68.94 _{-0.38}^{+0.39}$& $69.84$ & $70.00 _{-0.82}^{+0.42}$\\
$\beta_1$& $0.0$ & $<0.016$& $0.039$ & $<0.141$& $0.016$ & $<0.073$& $0.08$ & $<0.127$\\
$\beta_2$& $-$ & $-$& $0.177$ & $<0.138$& $0.0$ & $<0.067$& $0.097$ & $<0.137$\\
$\beta_3$& $-$ & $-$& $0.018$ & $<0.089$& $0.026$ & $<0.058$& $0.008$ & $<0.100$\\
$\beta_4$& $-$ & $-$& $0.037$ & $<0.034$& $0.034$ & $<0.023$& $0.039$ & $<0.05$\\
$\beta_5$& $-$ & $-$& $0.017$ & $<0.034$& $0.003$ & $<0.028$& $0.040$ & $<0.051$\\
$\beta_6$& $-$ & $-$& $0.01$ & $<0.048$& $0.004$ & $<0.047$& $0.001$ & $<0.058$\\
$\beta_7$& $-$ & $-$& $0.062$ & $<0.074$& $0.017$ & $<0.031$& $0.124$ & $<0.108$\\
\hline
$\Delta \chi^2$ & $-7.98$ && $-14.06$ && $-17.84 $&&$-13.74$& \\
$\ln{B}$&&& $-10.56$ && &&&\\
\hline 
 \end{tabular}   
 \caption{Table of best-fit, mean and 68\% C.L.  values of the various cosmological parameters of the 7-bin tomographic model, but using the baseline CMB dataset with a cut in Planck, i.e. Planck650+ACT+SPT. In the last two rows, we include the $\Delta \chi^2 = \chi^2_\mathrm{min,CDE}-\chi^2_\mathrm{min,\Lambda CDM}$ difference for each model (constant and tomographic CDE) and dataset, compared with $\Lambda$CDM using the same dataset, as well as the logarithm of the Bayes' ratio $\ln{B}=\ln{E_\mathrm{CDE}}-\ln{E_{\Lambda \mathrm{CDM}}}$ for one chosen dataset.}
 \label{table:planckcut_bestfit}
\end{table*}

\begin{table*}
\centering
\begin{tabular}{| *{7}{c|} }
 \hline

                    & \multicolumn{2}{c|}{\textbf{Cosmic Shear}}
            & \multicolumn{2}{c|}{\textbf{Galaxy Clustering$_s$}}
            & \multicolumn{2}{c|}{\textbf{3x2pt}}

                    \\
                &\multicolumn{2}{c|}{\textbf{($\Lambda$CDM)}}
                   &\multicolumn{2}{c|}{\textbf{($\Lambda$CDM)}} &\multicolumn{2}{c|}{\textbf{($\Lambda$CDM)}}
\\
                    
                    \hline
Parameter & Best-Fit & Mean$\pm\sigma$& Best-Fit & Mean$\pm\sigma$& Best-Fit & Mean$\pm\sigma$\\ \hline 
$100~\omega_\mathrm{b}$& $2.052$ & $2.257 _{-0.220}^{+0.216}$& $2.249$ & $2.233 _{-0.224}^{+0.226}$& $2.433$ & $2.241 _{-0.225}^{+0.249}$\\
$\omega_\mathrm{cdm}$& $0.139$ & $0.152_{-0.055}^{+0.043}$& $0.129$ & $0.124 _{-0.010}^{+0.001}$& $0.136$ & $0.123\pm 0.013$\\
$n_\mathrm{s}$& $0.991$ & $0.974 _{-0.076}^{+0.075}$& $0.863$ & $0.896 _{-0.049}^{+0.026}$& $0.864$ & $0.912 _{-0.060}^{+0.049}$\\
$H_0$& $67.37$ & $72.94 _{-5.50}^{+5.60}$& $70.21$ & $69.48\pm 2.20$& $71.63$ & $69.30 _{-2.60}^{+2.70}$\\
$\sigma_8$& $0.709$ & $0.737 _{-0.141}^{+0.107}$& $0.724$ & $0.734 _{-0.042}^{+0.037}$& $0.755$ & $0.759 _{-0.025}^{+0.023}$\\
$S_8$& $0.770$ & $0.750 _{-0.022}^{+0.032}$& $0.734$ & $0.739 _{-0.041}^{+0.038}$& $0.772$ & $0.764 _{-0.014}^{+0.017}$\\
\hline 
            & \multicolumn{2}{c|}{\textbf{Cosmic Shear}}
            & \multicolumn{2}{c|}{\textbf{Galaxy Clustering$_s$}}
            & \multicolumn{2}{c|}{\textbf{3x2pt}}
                    \\
                                & \multicolumn{2}{c|}{\textbf{(constant $\beta$)}}
            & \multicolumn{2}{c|}{\textbf{(constant $\beta$)}}
            & \multicolumn{2}{c|}{\textbf{(constant $\beta$)}}
                    \\

\hline 
$100~\omega_\mathrm{b}$& $2.350$ & $2.436 _{-0.111}^{+0.151}$& $2.516$ & $2.448 \pm 0.049$& $2.521$ & $2.478 _{-0.033}^{+0.047}$\\
$\omega_\mathrm{cdm}$& $0.1314$ & $0.1385 _{-0.0389}^{+0.0275}$& $0.1242$ & $0.1207 _{-0.0054}^{+0.0050}$& $0.1205$ & $0.1196 _{-0.0055}^{+0.0065}$\\
$n_\mathrm{s}$& $0.968$ & $0.953 _{-0.06}^{+0.057}$& $0.997$ & $0.970 _{-0.021}^{+0.017}$& $0.965$ & $0.975 _{-0.008}^{+0.011}$\\
$H_0$& $73.47$ & $73.06\pm 5.7$& $71.02$ & $70.44 _{-1.2}^{+1.1}$& $71.75$ & $71.15 _{-1.3}^{+1.4}$\\
$\sigma_8$& $0.752$ & $0.753 _{-0.092}^{+0.091}$& $0.802$ & $0.763 _{-0.042}^{+0.037}$& $0.810$ & $0.795 _{-0.021}^{+0.020}$\\
$S_8$& $0.735$ & $0.748 _{-0.023}^{+0.026}$& $0.797$ & $0.753 _{-0.038}^{+0.034}$& $0.787$ & $0.775\pm 0.018$\\
$\beta$& $0.02$ & $<0.064$& $0.002$ & $<0.019$& $0.004$ & $0.015 _{-0.011}^{+0.007}$\\
\hline
$\Delta \chi^2$ & $-2.48 $ && $-3.64$ && $-3.56$ &\\
$\ln{B}$ & $-11.35 $ &&&& &\\
\hline 
            & \multicolumn{2}{c|}{\textbf{Cosmic Shear}}
            & \multicolumn{2}{c|}{\textbf{Galaxy Clustering$_s$}}
            & \multicolumn{2}{c|}{\textbf{3x2pt}}
                    \\
                                                    & \multicolumn{2}{c|}{\textbf{(4-bin $\beta$)}}
            & \multicolumn{2}{c|}{\textbf{(4-bin $\beta$)}}
            & \multicolumn{2}{c|}{\textbf{(4-bin $\beta$)}}
                    \\

\hline 
$100~\omega_\mathrm{b}$& $2.413$ & $2.440 _{-0.070}^{+0.104}$& $2.495$ & $2.459 _\pm 0.030$& $2.294$ & $2.249 _{-0.024}^{+0.023}$\\
$\omega_\mathrm{cdm}$& $0.1231$ & $0.1319 _{-0.0445}^{+0.0358}$& $0.1349$ & $0.1310 _{-0.0084}^{+0.0080}$& $0.1239$ & $0.1261 _{-0.0047}^{+0.0049}$\\
$n_\mathrm{s}$& $0.949$ & $0.981 _{-0.036}^{+0.029}$& $0.884$ & $0.902 _{-0.059}^{+0.025}$& $0.909$ & $0.899 \pm 0.008$\\
$H_0$& $71.88$ & $72.08 _{-5.6}^{+5.0}$& $71.96$ & $71.18\pm 1.5$& $70.56$ & $70.03\pm 1.1$\\
$\sigma_8$& $0.794$ & $0.753 _{-0.122}^{+0.111}$& $0.765$ & $0.772 _{-0.050}^{+0.048}$& $0.799$ & $0.772 _{-0.022}^{+0.020}$\\
$S_8$& $0.774$ & $0.737 _{-0.032}^{+0.036}$& $0.776$ & $0.781 _{-0.051}^{+0.049}$& $0.792$ & $0.775 _{-0.019}^{+0.016}$\\
$\beta_1$& $0.059$ & $<0.151$& $0.016$ & $<0.018$& $0.003$ & $0.014 _{-0.010}^{+0.007}$\\
$\beta_2$& $0.22$ & $<0.166$& $0.01$ & $<0.021$& $0.008$ & $0.018 _{-0.011}^{+0.007}$\\
$\beta_3$& $0.09$ & $<0.154$& $0.014$ & $<0.023$& $0.013$ & $0.016 _{-0.010}^{+0.007}$\\
$\beta_4$& $0.029$ & $<0.107$& $0.013$ &$0.016 _{-0.011}^{+0.007}$& $0.021$ & $0.015 _{-0.011}^{+0.006}$\\
\hline
$\Delta \chi^2$ & $-2.81$ && $-2.57$ &&$-3.54$ & \\
$\ln{B}$ & $-6.73 $ &&&& &\\
\hline
 \end{tabular}   
 \caption{Table of best-fit, mean and 68\% C.L. values of the various cosmological parameters, for weak lensing, spectroscopic galaxy clustering and 3x2pt datasets. We report the results obtained with the $\Lambda$CDM, the $\beta=$const. CDE model and the 4-bin tomographic model. For reference, the binning is defined by edges $z=\{0,0.5,1,2\}$. In the last two rows of each model, we include the $\Delta \chi^2 = \chi^2_\mathrm{min,CDE}-\chi^2_\mathrm{min,\Lambda CDM}$ difference for each model (constant and tomographic CDE) and dataset, compared with $\Lambda$CDM using the same dataset, as well as the logarithm of the Bayes' ratio $\ln{B}=\ln{E_\mathrm{CDE}}-\ln{E_{\Lambda \mathrm{CDM}}}$ for one chosen dataset.}
 \label{table:WL_bestfit}
\end{table*}

\begin{table*}
\centering
\begin{tabular}{| *{5}{c|} }
 \hline
              & \multicolumn{2}{c|}{\textbf{Planck+ACT1800} }
              & \multicolumn{2}{c|}{\textbf{Planck+ACT1800} }
\\
                    &\multicolumn{2}{c|}{\textbf{+SPT+BSC+CS }}
                     &\multicolumn{2}{c|}{\textbf{+SPT+BSC+CS }}

                    \\
            &\multicolumn{2}{c|}{\textbf{($\Lambda$CDM) }}
&\multicolumn{2}{c|}{\textbf{(4-bin $\beta$)}}

                    \\
                    \hline
Parameter & Best-Fit & Mean$\pm\sigma$& Best-Fit & Mean$\pm\sigma$\\ \hline
$100~\omega_\mathrm{b}$& $2.433$ & $2.276 _{-0.224}^{+0.244}$& $2.288$ & $2.326 _{-0.179}^{+0.224}$\\
$\omega_\mathrm{cdm}$& $0.1492$ & $0.1617 _{-0.057}^{+0.0562}$& $0.1194$ & $0.1194\pm 0.0008$\\
$n_\mathrm{s}$& $0.925$ & $0.971 _{-0.081}^{+0.084}$& $0.974$ & $0.978\pm 0.003$\\
$H_0$& $71.37$ & $73.3 _{-6.0}^{+5.8}$& $69.55$ & $69.66 _{-0.36}^{+0.34}$\\
$\sigma_8$& $0.727$ & $0.72 _{-0.145}^{+0.130}$& $0.831$ & $0.831\pm 0.007$\\
$S_8$& $0.774$ & $0.744 _{-0.030}^{+0.034}$& $0.823$ & $0.822 _{-0.009}^{+0.010}$\\
$\beta_1$&$-$&$-$& $0.057$ & $0.056_{-0.052}^{+0.022}$\\
$\beta_2$&$-$&$-$& $0.154$ & $<0.092$\\
$\beta_3$&$-$&$-$& $0.033$ & $0.037_{-0.024}^{+0.017}$\\
$\beta_4$&$-$&$-$& $0.027$ & $<0.032$\\
\hline
$\Delta \chi^2$&&& $ -7.70$& \\
$\ln{B}$ &&& $-2.48 $&\\
\hline 
 \end{tabular}   
 \caption{Table of best-fit, mean and 68\% C.L. values of the various cosmological parameters, obtained with the dataset Planck+ACT1800+SPT+BSC+CS, for both a $\Lambda$CDM and tomographic CDE model. For reference, the binning is defined by edges $z=\{0,0.5,1,2\}$. In the last two rows, we show the value of $\Delta \chi^2 = \chi^2_\mathrm{min,CDE}-\chi^2_\mathrm{min,\Lambda CDM}$, as well as the logarithm of the Bayes' ratio $\ln{B}=\ln{E_\mathrm{CDE}}-\ln{E_{\Lambda \mathrm{CDM}}}$.}
 \label{table:planck_WL_bestfit}
\end{table*}

\section{Additional Data Probes}\label{sec:AppendixB}

Here we show results obtained from additional data probes that we have considered.
\subsection{3-bin Tomographic $\beta$ with Planck+PlanckLens}\label{sec:AppendixB1}
\begin{figure}[h]
    \centering
    \includegraphics[width=\linewidth]{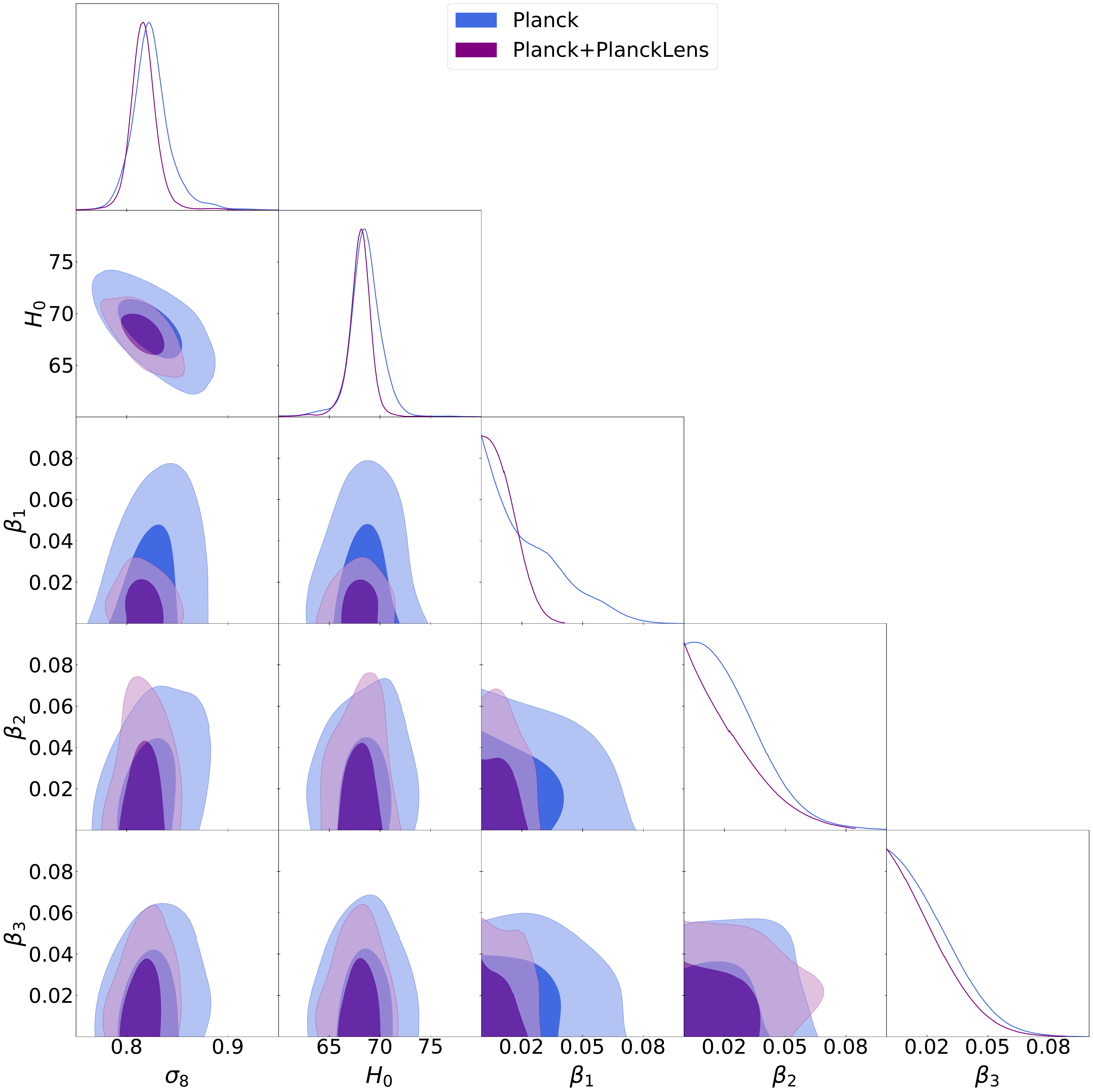}
    \caption{Comparison between the results obtained for the 3-bin tomographic model using Planck with (purple) and without (blue) the Planck CMB lensing likelihood. We recall that all spectra (in both cases) are always lensed, even when the Planck lensing (4-point) likelihood is not included.}
    \label{fig:lensing}
\end{figure}

We report results derived from additionally employing the $\textit{Planck}$ lensing likelihood in Fig.~\ref{fig:lensing}. We see that the inclusion of the lensing likelihood leads to tighter constraints, especially for the coupling at the lowest redshift bin, $\beta_1$. This is to be expected, since the CMB lensing potential power spectrum $C_\ell^{\phi\phi}$ probes the large scale structure at redshifts in that range, and has also been shown to lend a preference towards $\Lambda$CDM \cite{Planck:2018lbu}.

\subsection{7-bin Tomographic $\beta$ with Planck650+ACT+SPT cut}\label{sec:AppendixB2}
\begin{figure*}[ht]
    \centering
    \includegraphics[width=\linewidth]{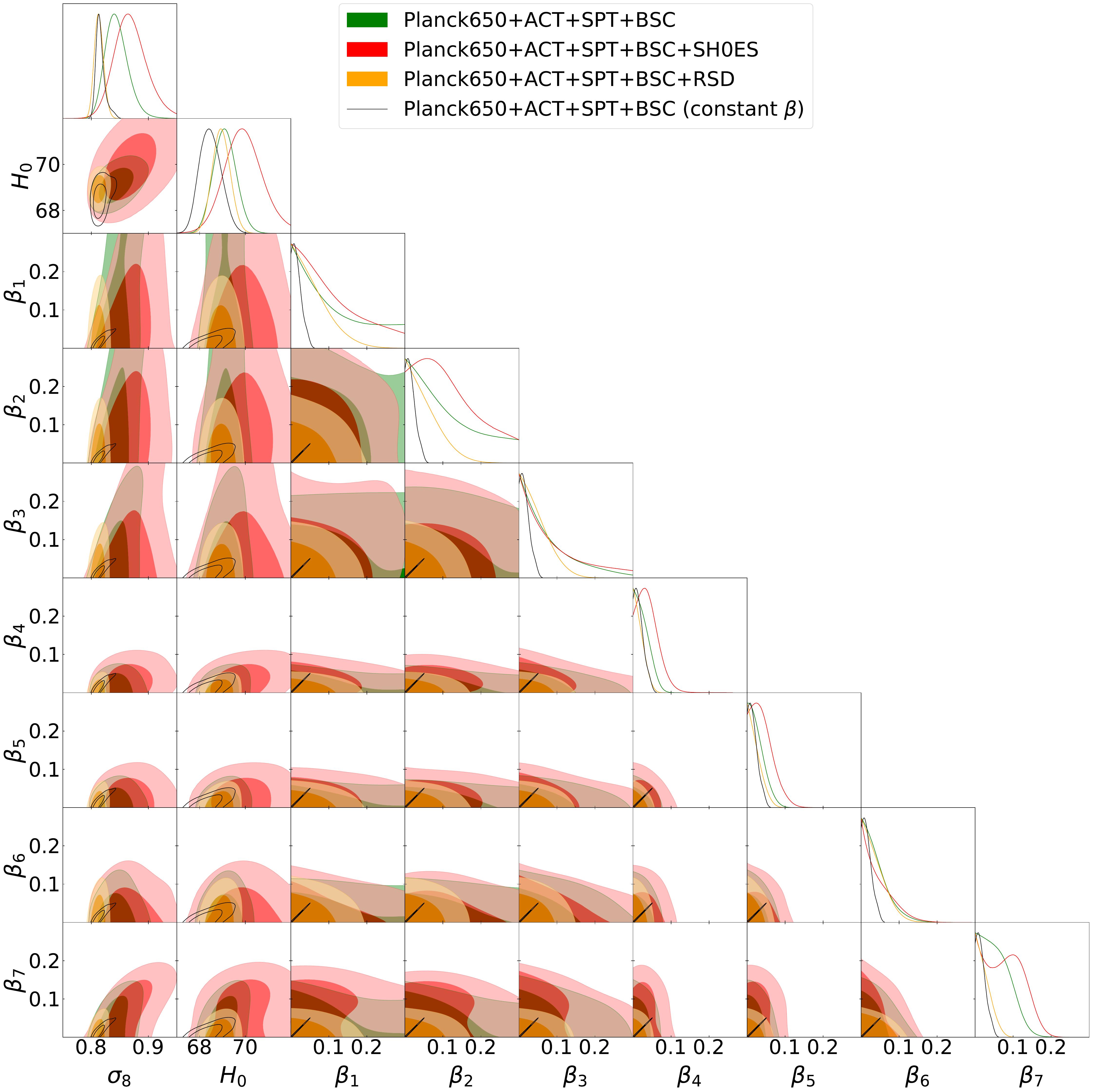}
    \caption{Triangular plot of 68\% and 95\% C.L. posterior distributions of $\sigma_8$, $H_0$, and the 7 tomographic coupling coefficients $\beta_{1-7}$ derived from datasets Planck650+ACT+SPT+BSC (green), Planck650+ACT+SPT+BSC+RSD (yellow) and Planck650+ACT+SPT+BSC+SH0ES (red). For reference, the binning is defined by edges $z=\{0,1,2,5,100,500,1000\}$. We include, in black lines, the contours obtained for a constant $\beta$ case with Planck650+ACT+SPT+BSC data. In this case, the contours for all $\beta_{1-7}$ are the same.}
    \label{fig:planckcut_full}
\end{figure*}

In Fig.~\ref{fig:planckcut_full}, we also present the results obtained with the baseline CMB dataset of Planck650+ACT+SPT, which corresponds, as explained above, to a larger cut of Planck data.  We report the best-fit, mean and 68\% C.L. uncertainties for each parameter in Table~\ref{table:planckcut_bestfit} of Appendix \ref{sec:AppendixA}. We see that when we add BSC we are unable to constrain the coupling strength at redshifts $z < 5$ (i.e. coupling parameters $\beta_1$, $\beta_2$ and $\beta_3$), similar to the case with Planck+ACT1800+SPT CMB dataset; but when RSD data is included,  constraints become tighter in all tomographic bins, and a lower value of $\sigma_8$ is favoured. Also, comparing with the contours obtained for a constant coupling case, and for the same dataset Planck650+ACT+SPT+BSC, in the 7-bin tomographic model, we no longer see any correlation between $\sigma_8$ and the coupling coefficient.

We also see that contours derived with the Planck650$+$ ACT$+$SPT$+$SH0ES dataset are much less constrained than that of the Planck+ACT1800+SPT+SH0ES dataset, with an interesting point that in the case of the former dataset, there is no significant evidence for a non-null coupling in any redshift bin. A cut on large $\textit{Planck}$ TT multipoles has shown to lend a preference for a higher value of $H_0$ already in the context of the $\Lambda$CDM \cite{Planck:2018vyg}, thus making the Planck650+SH0ES combination less in tension with SH0ES. This could explain why a nonzero coupling becomes less necessary. Furthermore, comparing between datasets, we see that a change between the Planck650+ACT+SPT and Planck+ACT1800+SPT dataset mainly affects coupling constraints at high redshifts, with the Planck650 cut allowing for larger values of $\beta_7$. This is analogous to the 3-bin model, where $\beta_3$, the amplitude of the bin at the highest redshift $z > 1000$, is the least constrained when the Planck650+ACT+SPT dataset is used. 

\begin{figure}[t!]
    \centering
    \includegraphics[width=\linewidth]{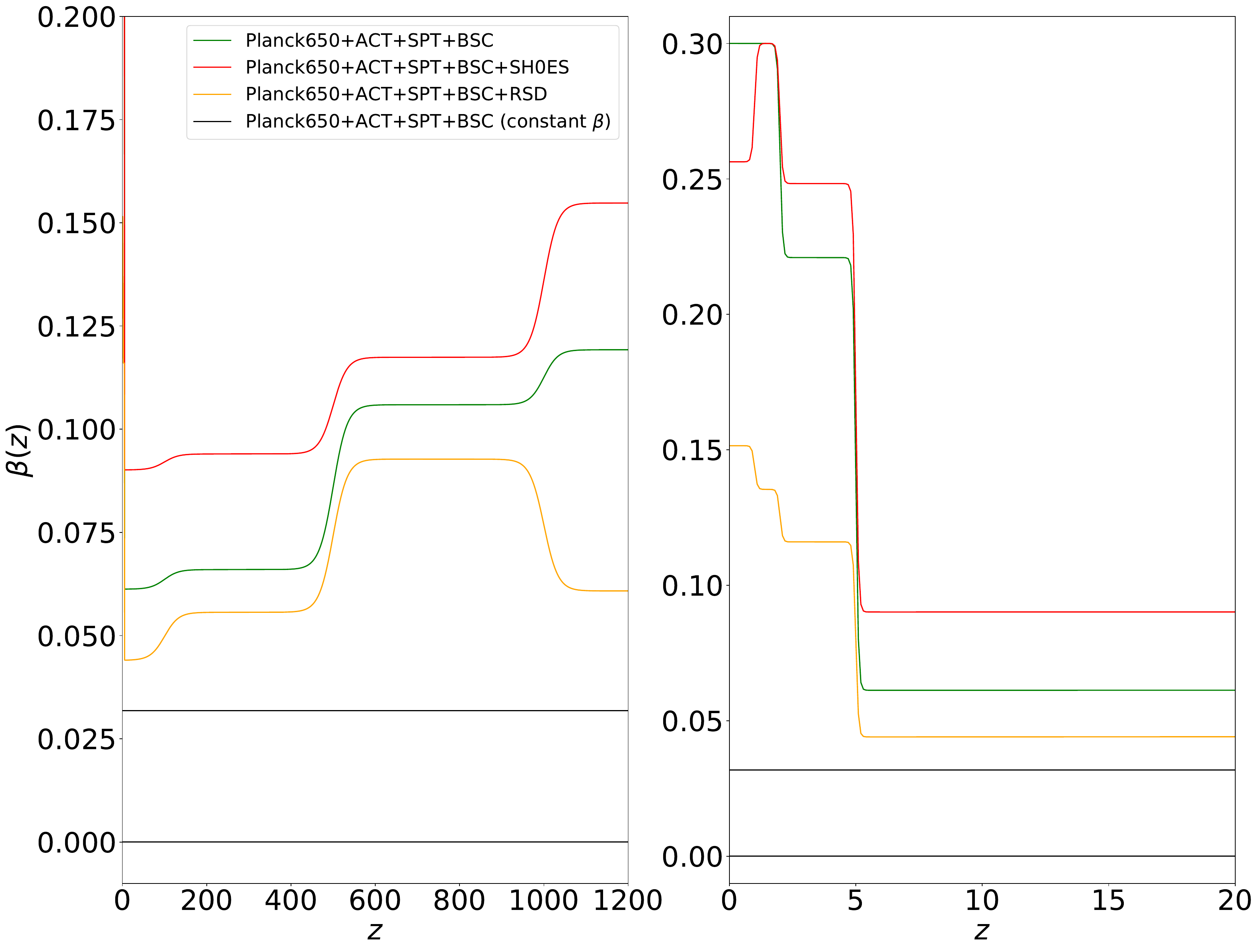}
    \caption{95\% C.L. on $\beta(z)$ for a 7-bin coupling model, with base CMB dataset Planck650+ACT+SPT. The right plot is a low-redshift zoom of the plot on the left.  For reference, we include the constraints for $\beta$ in the case of a constant coupling derived from a dataset Planck650+ACT+SPT+BSC in black. The binning is defined by edges $z=\{0,1,2,5,100,500,1000\}$.}
    \label{fig:planckcut_beta_lims}
\end{figure}

In Fig.~\ref{fig:planckcut_beta_lims} we plot $\beta(z)$ as a function of redshift, where similar to Fig.~\ref{fig:actpolcut_beta_lims}, constraints for every tomographic coupling bin are considerably loosened compared to a constant coupling model, especially at high redshifts. However, in all 3 datasets, results for $\beta$ are still compatible with 0 at all redshifts, unlike with the CMB baseline Planck+ACT1800+SPT dataset.

For these 3 datasets, the most tightly constrained bin is once again $\beta_4$, where $5<z<100$. This can be explained by the fact that this redshift bin encompasses the era of structure formation, to which a DE-DM coupling would be most sensitive, since as seen from Eqs.~(\ref{eq:consrhoDM}) and (\ref{eq:densityContrastDM}) coupling significantly affects the amount of dark matter and the amplitude of clustering. We further illustrate this in Fig.~\ref{fig:7-bin-Pk}, where we plot the resultant $P(k)$ as we activate the coupling at each subsequent bin 1 to 7. We see that increasing $\beta_4$ and $\beta_5$ from 0 to 0.1 gives the largest increase in $P(k)$. On the other hand, at lower redshifts (within tomographic bins $\beta_1$, $\beta_2$ and $\beta_3$), the impact of coupling on the matter power spectrum is much less significant, as seen from the inset of the same figure, which shows how activating $\beta_1$, $\beta_2$ and $\beta_3$ leads to only a slight increase in $P(k)$. Hence this could be why $\beta_{1-3}$ can take on larger values and are thus much less constrained. 

\bibliographystyle{apsrev4-1}
\bibliography{main}

\end{document}